\documentclass[lettersize,journal]{IEEEtran}
\usepackage{graphicx}
\usepackage{amsmath}
\usepackage{multirow}
\usepackage{colortbl}
\usepackage{color}
\usepackage{soul}
\usepackage{bm}
\usepackage{enumerate}
\usepackage{bm}
\usepackage[caption=false]{subfig}
\usepackage[table]{xcolor}
\usepackage{pifont}
\usepackage{paralist}
\usepackage{siunitx} 
\usepackage{array}   
\usepackage{multirow} 

\usepackage{amsthm}
\usepackage{siunitx}
\usepackage{atbegshi} 
\usepackage{siunitx}
\usepackage{flushend}
\usepackage{amssymb}
\usepackage{paralist}
\usepackage{siunitx}
\usepackage{epstopdf}
\usepackage[inline]{enumitem}

\usepackage{subcaption}
\usepackage{algorithm}
\usepackage{float}
\usepackage{tabularx}
\usepackage{algpseudocode}
\IEEEoverridecommandlockouts

\usepackage{parskip}
\usepackage{textcomp}
\usepackage{gensymb}
\usepackage{stfloats}
\usepackage{tikz}
\usepackage{verbatim}
\usetikzlibrary{chains,positioning,shapes.symbols,arrows,shapes,shadows,trees}
\usepackage{standalone}
\usepackage{array}
\usepackage{cite}
\usepackage{cuted, nccmath}
\usepackage{lipsum}
\usepackage{cleveref}
\usepackage{comment}
\usepackage[utf8]{inputenc}
\usepackage[T1]{fontenc}


\makeatletter
\newcommand{\multiline}[1]{%
  \begin{tabularx}{\dimexpr\linewidth-\ALG@thistlm}[t]{@{}X@{}}
    #1
  \end{tabularx}
}
\makeatother

 \renewcommand{\arraystretch}{1.5}

\algnewcommand{\AND}{\wedge}

\begin{document}

\setlength{\parindent}{1em}
\setlength{\parskip}{0pt}
\bstctlcite{IEEEexample:BSTcontrol}

\title{Molecular Arithmetic Coding (MoAC) and  Optimized Molecular Prefix Coding (MoPC) for Diffusion-Based Molecular Communication}

\author{Melih~Şahin,~\IEEEmembership{Student Member,~IEEE}, 
  Beyza~E.~Ortlek,~\IEEEmembership{Student Member,~IEEE}, Ozgur~B.~Akan,~\IEEEmembership{Fellow,~IEEE}
  \thanks{\hspace{2mm} Melih Şahin, Beyza E. Ortlek, and Ozgur B. Akan are with the Center for neXt-generation Communications (CXC), Department of Electrical and Electronics Engineering, Koç University, Istanbul 34450, Türkiye (e-mail: \{melihsahin21, bortlek14, akan\}@ku.edu.tr).}

		\thanks{\hspace{2mm} Ozgur B. Akan is also with the Internet of Everything (IoE) Group,	Department of Engineering, University of Cambridge, Cambridge CB3 0FA, U.K. (e-mail: oba21@cam.ac.uk).}

		\thanks{This work was supported in part by the AXA Research Fund (AXA Chair
for Internet of Everything at Koç University).}
}

\maketitle
\thispagestyle{empty}

\begin{abstract}


\\
Molecular communication (MC) enables information transfer through molecules at the nano-scale. This paper presents new and optimized source coding (data compression) methods for MC. In a recent paper, prefix source coding was introduced into the field, through an MC-adapted version of the Huffman coding. We first show that while MC-adapted Huffman coding improves symbol error rate (SER), it does not always produce an optimal prefix codebook in terms of coding length and power. To address this, we propose optimal molecular prefix coding (MoPC). The major result of this paper is the  Molecular Arithmetic Coding (MoAC), which we derive based on an existing general construction principle for constrained arithmetic channel coding, equipping it with error correction and data compression capabilities for any finite source alphabet. We  theoretically and practically show the superiority of MoAC to SAC, our another  adaptation of arithmetic source coding to MC. However, MoAC's unique decodability is limited by bit precision. Accordingly, a uniquely-decodable new coding scheme named Molecular Arithmetic with Prefix Coding (MoAPC) is introduced. On two nucleotide alphabets, we show that MoAPC has a better compression performance than MoPC. MC simulation results demonstrate the effectiveness of the proposed methods.
\end{abstract}

\vspace{0.1cm}
\vspace{-1cm}

\vspace{-0.1cm}

\IEEEpeerreviewmaketitle

\vspace{1.7cm}
\section{Introduction}
\vspace{-1pt}

\IEEEPARstart{M}{olecular} communication (MC) is a bio-inspired communication method aiming to  transmit information in the characteristics of chemical signals. These signals are released and detected by molecular entities, such as cells or nano-scale devices. The simplest form of MC involves diffusion, where each molecule moves pseudo-randomly through space via Brownian Motion\cite{nakano_eckford_haraguchi_2013,ozgur_hoca_survey}. In the communication model, as depicted in Fig. \ref{fig1}, the transmitter releases a number of specific class of molecules. The receiver then attempts to decode the signal based on the detection timings, the quantity, and the types of the molecules it identifies \cite{MC_main}. However, this kind of communication leads to a phenomenon known as inter-symbol-interference (ISI), which occurs when the information molecules transmitted in the communication system overlap or interfere with each other in subsequent signal intervals, causing the communication channel to have a high memory component \cite{ISI,ISI_mitigating_methods_2015}.

\setlength{\floatsep}{0pt}
\setlength{\textfloatsep}{0pt}
\setlength{\intextsep}{0pt}
\setlength{\abovecaptionskip}{3pt}
\setlength{\belowcaptionskip}{3pt}

Source coding (data compression) methods reduce the number of bits needed to encode the data \cite{survey_of_bio_informatics}. For MC via diffusion, reducing the number of bits required for lossless data transmission can improve channel quality by allowing for longer signal intervals and thus reduce ISI. Conversely, channel coding introduces additional bits for data redundancy, which can shorten signal intervals and increase ISI. However, certain channel coding techniques have error-correcting properties that can outweigh the disadvantage of shorter signal intervals, as demonstrated in \cite{best_channel_coding_2020}. Integrating source coding with channel coding strategies leverages the strengths of both approaches. A recent study supports the benefits of this integrated approach: Simulations in \cite{lee2023isimitigating} demonstrated that integrating Huffman source coding \cite{huffman_original_paper} with ISI-mitigating channel codes \cite{best_channel_coding_2020} results in significant improvements in symbol error rate (SER) compared to using Huffman coding alone. 

Arithmetic coding \cite{arithmetic_coding} is known to be better than Huffman coding in terms of its data compression rate \cite{data_compression_book} \cite{better}. Additionally, arithmetic coding is utilised in many of the most effective biological data compression algorithms \cite{survey_of_bio_informatics}. Considering MC will mostly find application areas in biological organisms \cite{nakano_eckford_haraguchi_2013}, the adaption of arithmetic source coding, of which use is highly prevalent in bio-informatics, into MC is necessary. Accordingly, in this paper, we develop two novel arithmetic source coding methods: SAC and MoAC.
 
 Our paper is organized as follows: Section \ref{sec:second}.A defines the system model. The remaining subsections of Section \ref{sec:second} presents our proposed coding schemes: optimized molecular prefix coding (MoPC$^*$), substitution arithmetic coding (SAC), molecular arithmetic coding (MoAC), and molecular arithmetic with prefix coding (MoAPC). Section \ref{detection} discusses detection and error correction algorithms. Section \ref{sec:third} provides a comparative analysis of the proposed source coding methods.

\vspace{-2pt}
\section{Arithmetic and Prefix Source Coding for Molecular Communication}
\vspace{2pt}
\label{sec:second}
\vspace{-0.05cm}
\subsection{System Model}
\vspace{-0.05cm}

This paper assumes a molecular communication via diffusion (MC) channel, where a point transmitter releases a predetermined number of information molecules into the environment at the start of each signal interval with a constant symbol duration \({t_{s}}\). These information molecules move in a pseudo-random manner through the 3-dimensional fluidic environment, following the principles of Brownian motion as described in \cite{nakano_eckford_haraguchi_2013}. The receiver absorbs any information molecule that comes into contact with its surface and keeps track of the count of these molecules for each signal interval. 

\begin{figure}[t]
\centering
\includegraphics[width=0.9\linewidth]{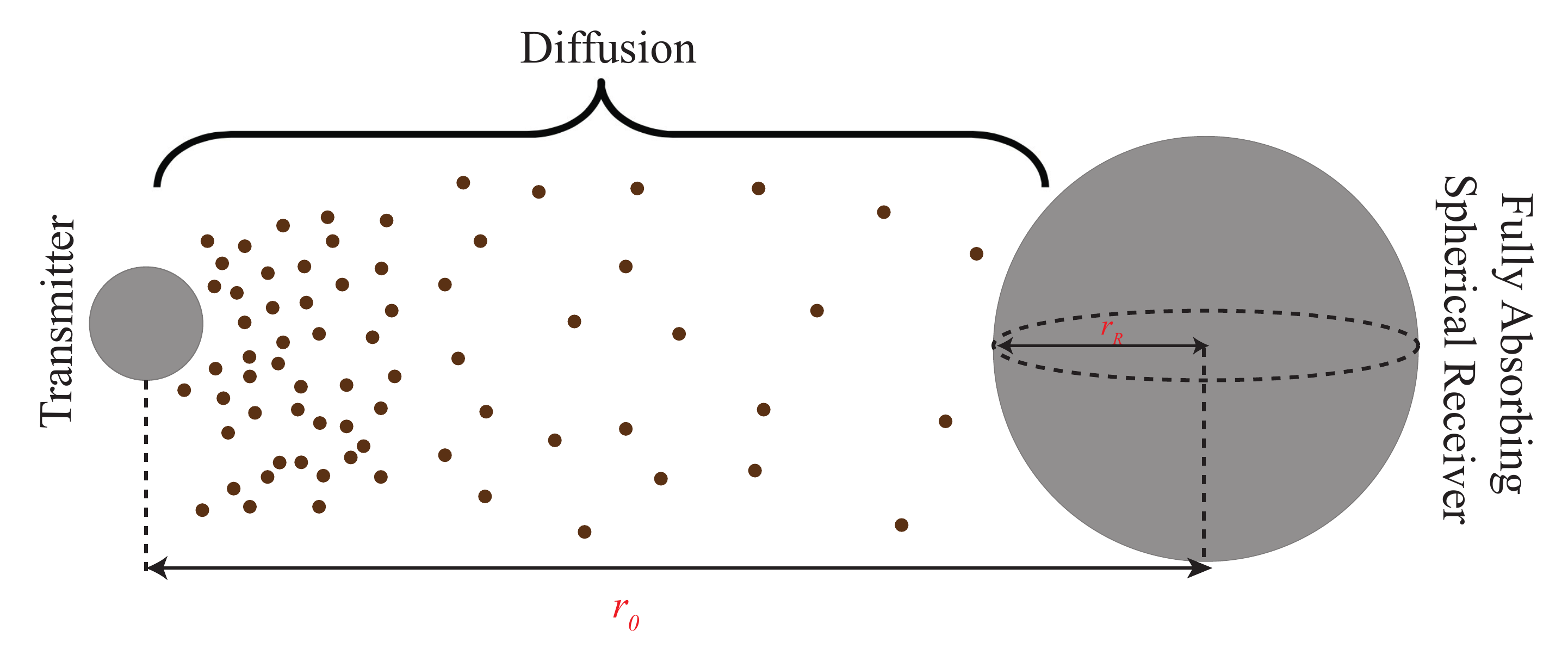}
\caption{MC Channel}
\label{fig1}
\end{figure}

This process is depicted in Fig. \ref{fig1}, where \( r_{\scriptscriptstyle R} \) is the radius of the spherical receiver, \( r_{0} \) is the distance between the center of the spherical receiver and the point transmitter. At each time step, $\Delta t$ (in seconds), the position of a  molecule $(x,y,z)$ is updated along each coordinate axis as follows \cite{diffusion_MC_sim,diffusion_normal_distribution}:
\vspace{-3pt}
\begin{gather*}
\Delta x,\Delta y,\Delta z \sim \mathcal{N}(0, 2\cdot D\cdot \Delta t ), \tag{1}
\end{gather*}
\vspace{-0.55cm}
\vspace{-3pt}

\hspace{-0.4cm}where $D$ is the diffusion coefficient. We modulate the information through the quantity of the  information molecules emitted at the start of each signal interval. If the current signal interval corresponds to a 1-bit, transmitter emits a pre-defined number of information molecules. For a 0-bit, the transmitter does not emit any information molecule. In MC, this is known as the binary concentration shift keying (BCSK) \cite{BCSK}. 
\vspace{5pt}

\vspace{-0.6cm}
\subsection{Optimized Molecular Prefix Coding (MoPC)}
\label{sec:sectionprefixcoding}

In the context of source coding, assigning each symbol with a code in such a way that none of the codes is a prefix of another code ensures unique decodability. For codebooks under no restriction, one technique to find a length-wise optimal prefix codebook is the Huffman coding \cite{huffman_original_paper}.

Authors of \cite{lee2023isimitigating}  combine source coding with the error correction properties of one of the most successful MC channel codes \cite{best_channel_coding_2020} by not allowing consecutive 1-bits in each resultant Huffman code through substituting each 1-bit with a 10. In this paper, we abbreviate the MC-adapted Huffman coding \cite{lee2023isimitigating} as MoHuffman. Though MoHuffman highly improves the symbol and word error rate values compared to Huffman coding, it does not always produce a length-wise optimal codebook. Consider the alphabet $(a,b,c,d,e)$ with respective probabilities $(0.201,0.201,0.201,0.199,0.198)$. MoHuffman produces a prefix code space $ a\to 00,b\to 010,c\to 1010, d\to 1000, e\to 10010$, with expected code length $3.595$. However, $ a\to 000,b\to 010,c\to 100, d\to 0010, e\to 1010$ is a better prefix codebook with expected code length $3.397$.

In literature, a prefix codebook whose codes avoid consecutive 1-bits and end with a 0-bit (or equivalently start with a 0-bit) is known as a $(1,\infty)$ constrained Huffman coding \cite{constrained_huf1,constrained_huf2}. In this paper, we will refer to it as Molecular Prefix Coding (MoPC), as MoPC also incorporates the ISI-mitigating error correction technique \cite{best_channel_coding_2020} (given in Algorithm 2 in Section III) whose usage for MC prefix coding is proposed in MoHuffman \cite{lee2023isimitigating}. Furthermore, MoPC also has an optimization constraint on transmission of 1-bits as will be clarified in this subsection. Finding a length-wise optimal MoPC is equivalent to the problem of constructing optimal prefix-free codes with unequal letter cost, as the cost of 1-bits and 0-bits can be taken as 2 and 1 respectively. Polynomial algorithms for this problem exist, with time complexity as little as $\mathcal{O}(n^2)$, where $n$ is the symbol alphabet size \cite{n2complexity}, \cite{poly}.

However, minimizing the coding length is not the only criterion; reducing the transmission of 1-bits is also important. In this paper, we first select the prefix codebook with the shortest length. If there are multiple length-wise optimal codebooks, we then choose the one that produces the lowest expected number of 1-bits. We abbreviate an optimal MoPC constructed under these conditions as MoPC$^{*}$. For instance, when given an alphabet $(a,b,c)$ with respective probabilities $(0.4,0.3,0.3)$, the length-wise optimal MoPC codebooks $ a\to 10,b\to 010,c \to 00 $ and $ a\to 00,b\to 010,c \to 10 $ both produce the same expected length, $0.4\cdot2+ 0.3\cdot3 +0.3\cdot2=2.3$. However, the first codebook has an  average per-symbol 1-bit transmission of $0.7$, while the second codebook has an average per-symbol 1-bit transmission of $0.6$. Therefore, since the second codebook would consume fewer information molecules, it should be preferred over the first one; and it is actually a MoPC$^{*}$.

To the best of our knowledge, there is not a polynomial algorithm to perform this task of choosing the codebook with the least average number of 1-bits among the length-wise optimal MoPC codebooks. Accordingly, MoPC$^{*}$ codebooks in the performance evaluation section of this paper have all been derived by a trivial brute-force algorithm, which searches through the space of all possible MoPC codebooks. 
\vspace{2pt}
\vspace{-10pt}
\subsection{Classical Arithmetic Coding (AC)}
\label{sec:classicarithmeticcoding}

\begin{figure}[t]

\centering
\includegraphics[width=0.59\linewidth]{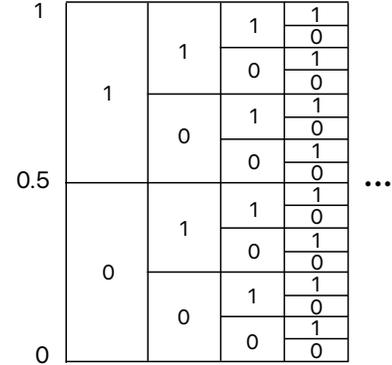}
\caption{Code Space Depiction of AC}

\end{figure}
\vspace{-2pt} 
This section presents an accessible introduction to classical arithmetic coding (AC), as its definitions and underlying logic will be frequently referenced throughout the subsequent sections. Each AC code (i.e., bit sequence) in the code space has a unique interval associated with it. For instance,  `1' is associated with   $[0.5,1)$; and `01' is associated with $[0.25,0.5)$ as in Fig. 2. Let $b_1 b_2 ... b_n$ be a code of length $n$. Then its corresponding interval, $[k, l)$ is defined as follows \cite{arithmetic_coding, data_compression_book}:

{\setlength{\abovedisplayskip}{1.5pt}
\setlength{\belowdisplayskip}{1.5pt}
\setlength{\abovedisplayshortskip}{1.5pt}
\setlength{\belowdisplayshortskip}{1.5pt}}
\vspace{-5pt}
\begin{equation}
[\sum_{k=1}^{n} b_k \cdot 2^{-k}, ~~2^{-n} + \sum_{k=1}^{n} b_k \cdot 2^{-k})
\label{eq:interval}
\tag{2}
\end{equation}

\begin{figure}[t]

\centering
\includegraphics[width=0.55\linewidth]{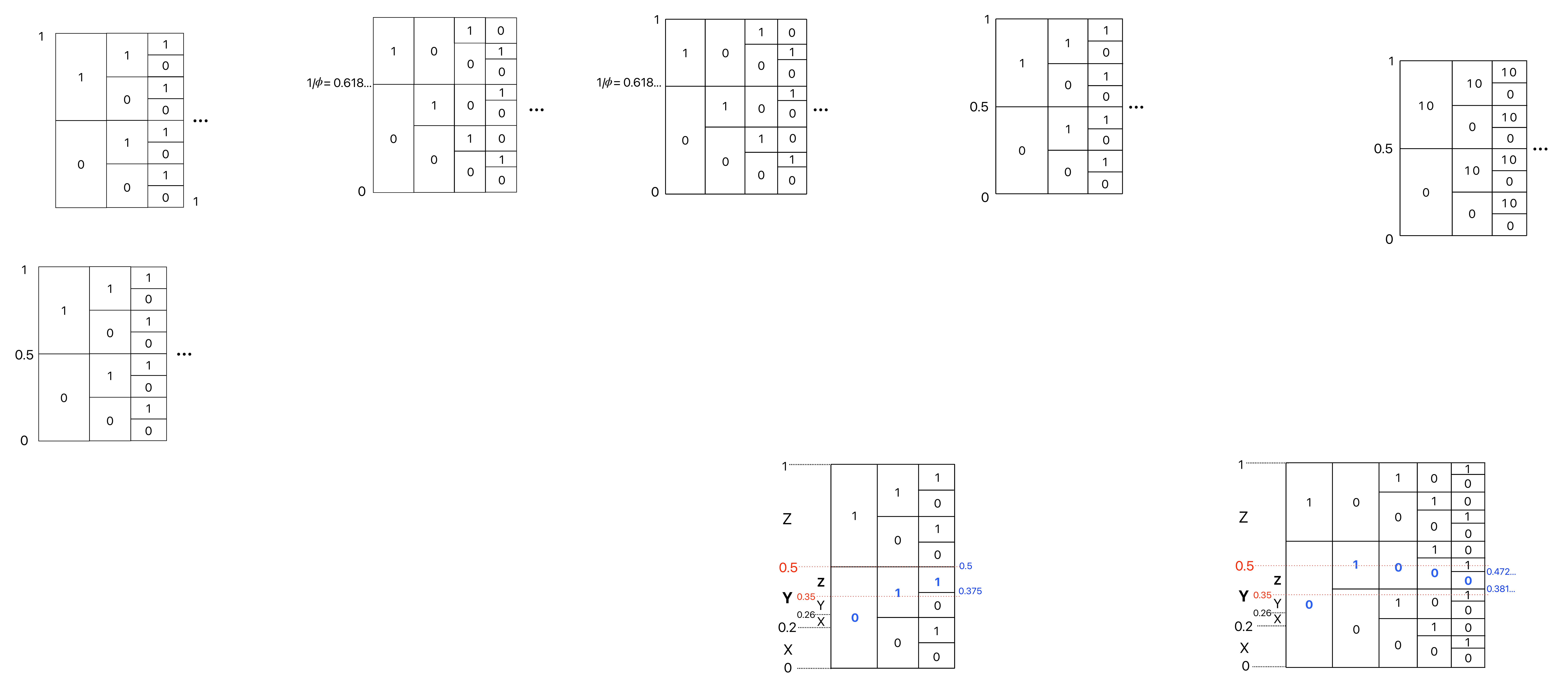}
\caption{AC Encoding of the Exemplary Word $YZ$}

\vspace{0.05cm}
\end{figure}

\vspace{-5pt}
When given a finite alphabet, together with the probabilities of each symbol, a unique interval $[a,b) \subseteq [0,1)$ can be associated with each word \cite{data_compression_book}. Let $word$ be a symbol sequence with length $n$ such that $word(i)$ denotes the $i^\text{th}$ symbol of the $word$, where $1\le i \le n$. Let our $alphabet$ set, which includes $N$ symbols be represented by a bijective ordering function $ord: alphabet \to \{1,...,N\}$. This way, each symbol of $alphabet$ can be associated with a number between $1$ and $N$.  Let $prob: \{1,...,N\} \to [0,1]$ be a function which maps each number $k$ to the probability of the symbol associated with the number $k$. For \( i = 1, \ldots, N \), define:
\vspace{-7pt}
\vspace{2pt}
\begin{equation}
c(i) = \sum_{k=1}^{i-1} \text{prob}(k), \quad d(i) = \sum_{k=1}^{i} \text{prob}(k)
\label{eq:c_d_definition}
\tag{3}
\end{equation}
\vspace{-0.2cm}
\vspace{-2pt}

The interval $[a,b) $ associated with $word$ can be recursively obtained using composition of functions, as follows \cite{arithmetic_coding}.

\begin{equation}
\begin{aligned}
&\text{Let } a_1 = c(ord(\text{$word$}(1))) \text{ and } b_1 = d(ord(\text{$word$}(1))). \\
&\text{For all } 2 \le i \le n, \text{ let } \\
&a_i = a_{i-1} + \left( c(ord(\text{$word$}(i))) \right) \cdot \left( b_{i-1} - a_{i-1} \right) \quad \\
&b_i = a_{i-1} + \left( d(ord(\text{$word$}(i))) \right) \cdot \left( b_{i-1} - a_{i-1} \right). \\
& \text{Define } [a,b)=[a_n, b_n).
\end{aligned}
\tag{4}
\vspace{1pt}
\end{equation}

Through using this unique interval $[a,b)$, a bit sequence can be associated with the given $word$. This bit sequence is the shortest code, having the corresponding interval, $[k,l)$, computed using (2), conforming $[k,l)\subseteq [a,b)$. If  the information of the length of the bit sequence is provided to the decoder, an end-of-file (EOF) symbol at the end of each word is not needed \cite{data_compression_book}.

This encoding procedure will now be illustrated through a simple example. Consider an exemplary EOF-included alphabet $(X,Y,Z)$ with respective probabilities $(0.2,0.3,0.5)$, where $Z$ serves as the EOF symbol. Define the ordering function $ord: \{X, Y, Z\} \rightarrow \{1, 2, 3\}$ as being $ord(X)\hspace{-5pt}=\hspace{-5pt}1, \, ord(Y)=2 \, $, and $ ord(Z)=3$.
 Then, using (3) and (4), for the exemplary word $YZ$, the corresponding interval is calculated to be $[0.35,0.5)$. The shortest bit sequence, whose corresponding interval is a subset of $[0.35,0.5)$, is $011$ as shown in Fig. 3. The interval of $011$ is $[0.375,0.5)$ from (2), which satisfies $[0.375,0.5) \subseteq [0.35,0.5)$ as it should.

For EOF-included decoding, assume a bit sequence \(\mathbf{b}= b_1 b_2 \ldots b_n b_{n+1} b_{n+2} \ldots b_{n+h} \), which comprises appended encodings of ensuing words, is given. Let the interval of \(\mathbf{b}\) be \([k_1, l_1)\) computed based on the definition at (2). Let the encoding of the initial world be $b_1 b_2 ... b_n$ with its interval being $[k_2,l_2)$, computed using (2). The decoding of the initial word of this whole bit sequence, \(\mathbf{b} \),  is the longest word in which the only EOF character is at its end, and whose corresponding interval $[a,b)$ satisfies $[k_1,l_1) \subseteq [a,b)$. The decoder does not know the position of the initial word's final bit, $b_n$. Since an EOF character is available, this is not an issue. Because the interval of $b_1 b_2 ... b_n b_{n+1} b_{n+2} ... b_{n+h}$ is a subset of the interval of $b_1 b_2 ... b_n$ (i.e $[k_1,l_1)\subseteq [k_2,l_2)$ ). Therefore,  any decoding of the bit sequence $b_1 b_2 ... b_n \cup S$, where $S$ represents all possible bit sequences of varying lengths, and $\cup$ is the sequence appending operator, would all be decoded as the initial word whose encoding is $b_1 b_2 ... b_n$.

As an instance, for the exemplary alphabet given in the paragraph two before, the codes $1010\cup S$, where $S$ is any bit sequence, would all be decoded as $YZ$. For EOF-excluded decoding, the decoder must have the knowledge of where the initial word's encoding ends in the given whole bit sequence. Then the initial words is decoded as being the longest word whose  encoding identically matches the given initial portion of the whole encoding.

\begin{figure}[t]
\centering
\includegraphics[width=0.52\linewidth]{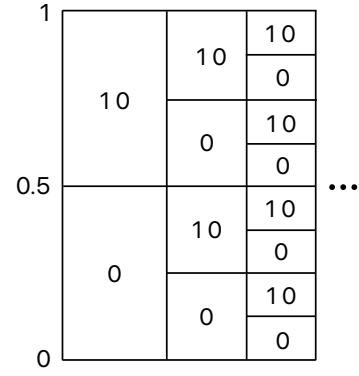}
\caption{Code Space Depiction of SAC}
\end{figure}

\vspace{-10pt}
\subsection{Substitution Arithmetic Coding (SAC)}
\label{sec:classicarithmeticcoding}

 Our purpose is to create an arithmetic coding method that ensures each 1-bit is followed by at least one 0-bit. This property is needed to achieve the error-correction property in an MC channel as proposed in \cite{best_channel_coding_2020}. One simple but inefficient solution, as we propose, is to substitute each 1-bit in the AC with a $10$. For instance the word $YZ$, using the encoding algorithm of the AC, is first encoded to be $011$. Then, in this new scheme, it would be converted to $01010$. Then to decode $01010$, we would substitute each $10$ with a 1-bit, converting it back to its original form, $011$. Then, using the decoder algorithm for the AC, the corresponding word $YZ$ would be found. We name this new scheme as the Substitution Arithmetic Coding (SAC). This adaption technique is very similar to the approach followed in MoHuffman \cite{lee2023isimitigating}. The code space depiction of SAC is given in Fig. 4.

\begin{figure}[t]
    \centering
    \includegraphics[width=0.7\linewidth]{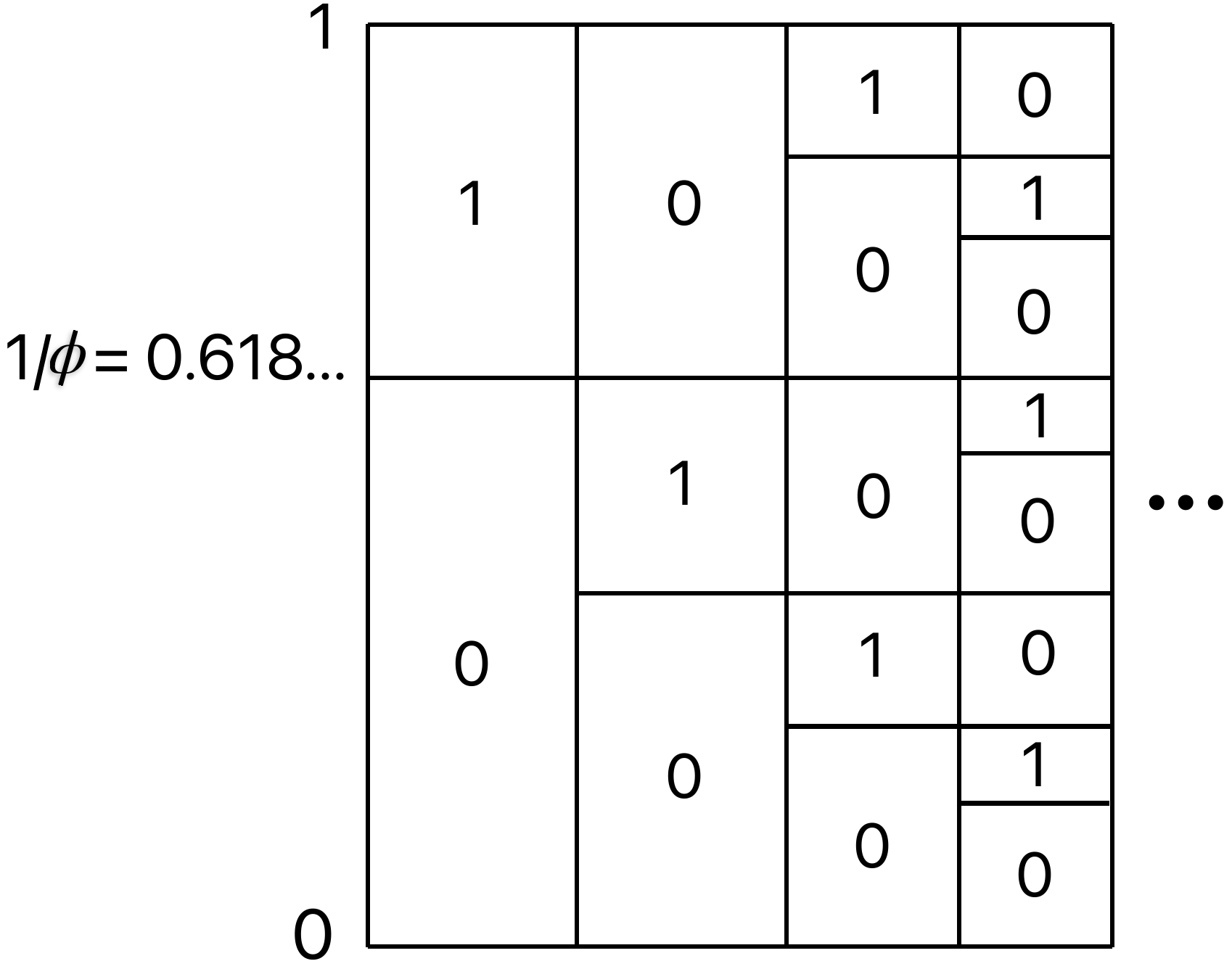}
    \vspace{0.1cm}
    \caption{ Code Space Depiction of MoAC}
    \label{fig3}  

\end{figure}

\vspace{-0.2cm}
\subsection{Molecular Arithmetic Coding (MoAC)}
\label{sec:classicarithmeticcoding}

In this section, building on the general scheme outlined in \cite{RLL_arithmetic}, we construct an arithmetic source coding method, which we name MoAC, with error correction capabilities achieved by avoiding consecutive 1-bits. We will demonstrate both theoretically and practically that MoAC offers substantial advantages in coding length and power consumption over SAC. The ISI-mitigating codes defined in \cite{best_channel_coding_2020}, without the existence of a 1-bit condition in each code, are equivalent to run-length-limited (RLL), also known as constrained, codes of order $(1,\infty)$. More generally, an RLL code of order $(d,k)$ enforces that a 1-bit must be followed by at least $d$ 0-bits, and no more than $k$ consecutive 0-bits can appear \cite{RLL}. The scheme in \cite{RLL_arithmetic} provides a method for constructing arithmetic codes for any RLL code of order $(d,k)$, specifically for channel coding. Additionally, in \cite{constrained_huf1}, this approach is extended to data compression for binary source alphabets. The proposed MoAC data compression scheme is designed to accommodate source alphabets of any finite size, extending beyond the binary case.

We will follow the general scheme in \cite{RLL_arithmetic}, applying it to $(1,\infty)$ RLL codes, with some modifications for MoAC.  Specifically, instead of adhering to the strict design where a 0-bit occupies the first column, MoAC removes this bit and requires each encoding to terminate with a 0-bit. This modification aligns MoAC with the structure of MoPC, proving useful for later defining MoAPC. Let $C(n)$ denote all $(1,\infty)$ RLL codes of length $n$ that do not require the existence of a 0-bit at the start of each code. First two terms of $C(n)$, which is known as the $1$-limited sequence \cite{RLL_without_condition}, are as follows:

\vspace{1pt}
\setlength{\abovedisplayskip}{0pt} 
\vspace{2pt}
\begin{equation}
C(1)=\{0,1\}, \quad C(2)=\{00,01,10\}
\tag{5}
\end{equation}
\vspace{-13pt}
\setlength{\abovedisplayskip}{0.2cm} 
\vspace{-1pt}

\vspace{-1pt}
To construct an arithmetic encoder for $C(n)$, growth rate $W$ of $C(n)$ needs to be derived \cite{RLL_arithmetic}. Let $|C(n)|$ denote the total number of bit sequences (i.e., codes) inside $C(n)$. From (5), $|C(1)|=2$ and $|C(2)|=3$. Note that $|C(n)|$ conforms to the recursive relation $|C(n)|=|C(n-1)|+|C(n-2)|$ \cite{RLL_without_condition}. This is because, all codes of $C(n)$ can be formed by inserting $0$ to the start of all $C(n-1)$, and by inserting $10$ to the start of all $C(n-2)$.  Thus $|C(n)|=Fibonacci[n+2]$. Then, $W$ for $C(n)$, using properties of $Fibonacci[n]$, is given by:

\begin{equation}
W=\lim_{{n \to \infty}} \frac{|C(n)|}{|C(n-1)|} = \phi = \frac{1 + \sqrt{5}}{2} = 1.618\ldots
\label{eq:golden_ratio}
\tag{6}
\end{equation}

\vspace{-3pt}
Note that, ($1,\infty$) RLL happens to have to the same W constant, $\phi$ \cite{constrained_huf2}. Using $\phi$, the code space structure of MoAC, as shown in Fig. \ref{fig3}, will now be defined. For $C(n)$ codes, a 1-bit must be followed by a 0-bit, while a 0-bit can be followed by either a 0-bit or 1-bit. Furthermore, the general scheme in  \cite{RLL_arithmetic} enforces a 1-bit to have an interval height of $(1/W)^{n+1}$, where $n$ is the column index. Please note that since we have not enforced the code space to begin with a 0-bit, we shifted the original formula of $(1/W)^{n}$ by $1$. Based on these rules, we recursively construct the MoAC code space as follows:

\begin{figure}[t]

\centering
\includegraphics[width=0.73\linewidth]{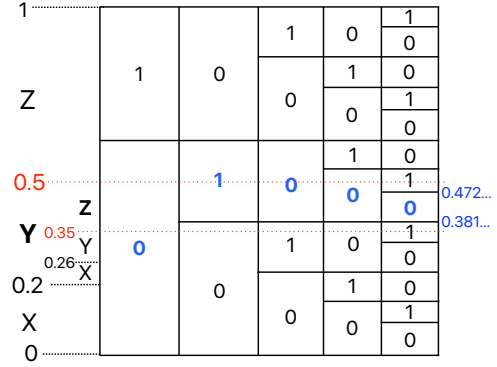}
\caption{MoAC Encoding of the Exemplary Word $YZ$}

\end{figure}

At the 1$^{st}$ column, the 0-bit is assigned the interval $[0, 1/\phi)$, and the 1-bit interval is assigned the interval $[1/\phi, 1)$ as shown in Fig 5. For any 1-bit in any $n^\text{th}$ column with interval assignment $[x, y)$, there is a 0-bit in the $(n+1)^\text{th}$ column with interval assignment $[x, y)$. For any 0-bit in any $n^\text{th}$ column with interval assignment $[x, y)$, there is a 0-bit in the $(n+1)^\text{th}$ column with interval assignment $[x, (x+(y\cdot\phi))/(1+\phi))$, and a 1-bit in the $(n+1)^\text{th}$ column with interval assignment $[(x+(y\cdot\phi))/(1+\phi), y)$. Let $b_1 b_2 ... b_n$ be a code of length $n$, which do not contain consecutive 1-bits. Then, we define its corresponding MoAC interval, $[k, l)$, as follows.

\vspace{-4pt}
\begin{equation}
[\sum_{i=1}^{n} b_i \cdot (1/\phi)^{-i},~ ~~(1/\phi)^{(-n+b_n) } + \sum_{i=1}^{n} (b_i \cdot (1/\phi)^{-i}))
\label{eq:interval}
\tag{7}
\end{equation}

{\setlength{\abovedisplayskip}{1.0pt}
\setlength{\belowdisplayskip}{1.0pt}
\setlength{\abovedisplayshortskip}{1.0pt}
\setlength{\belowdisplayshortskip}{1.0pt}}
\vspace{-4pt}
Using  (4), each word of a given alphabet can be associated with a unique MoAC interval, $[a,b)$. Then we define the MoAC encoding of a given word to be the shortest bit sequence that ends with a 0-bit and whose MoAC interval $[k,l)$, computed using (7), satisfies $[k,l)\subseteq [a,b)$. To illustrate the working mechanism of MoAC, we will use the same example given for AC. Let our exemplary EOF-included alphabet be $(X,Y,Z)$ with respective probabilities $(0.2,0.3,0.5)$. Let our exemplary word be $YZ$ as previously. Using (3) and (4), the interval of $YZ$ is computed to be $[0.35,0.5)$. Then the shortest MoAC sequence whose interval $[k,l)$ is a subset of $[0.35,0.5)$ is found to be $01000$ as illustrated in Fig. 6. The MoAC interval of  $01000$, using (7), is $[(1/\phi)^{2},(1/\phi)^{2}+(1/\phi)^{5})\approx[0.381,0.472)\subseteq[0.35,0.5)$.
\vspace{-0.5pt}
To decode \( 01000 \), we identify the shortest word where the EOF character (in this case, \( Z \)) exclusively appears at its end, and whose interval is a subset of \( 01000 \). This word is \( YZ \).

\vspace{-0.5pt}
Now it will be shown that MoAC produces shorter codes than SAC with an approximate ratio of 1 to 1.0413. In deriving this ratio, the following lemma will be used. Note that, if a code has an associated interval $[k, l)$, its interval height is defined to be $l-k$.

\vspace{-0.05cm}

\textit{Lemma:} The interval height values of MoAC codes having the same length $n$ and ending with a 0-bit are all equal. Similarly, all MoAC codes of length $n$ ending with a 1-bit have the same interval height value. For instance, MoAC codes $100$, $010$, and $000$ have identical interval height values.

\textit{Proof:}
    We proceed by induction. Initial induction statement for $n=1$ clearly holds, as at the first column of Fig. 3, there is only one 1-bit and 0-bit. As the inductive argument, assume that what is stated at the lemma above holds for an $i^\text{th}$ column, i.e. for all codes of length $i$. In the $i^\text{th}$ column, let the interval height of each 1-bit be $a$, and accordingly let the interval height of each 0-bit be $a\cdot\phi$. In the $(i+1)^\text{th}$ column, a 0-bit can come either after a 0-bit or a 1-bit. If it comes after a 1-bit, its length is same as the 1-bit, i.e. $a$. If it comes after a 0-bit, its length is $(a\cdot\phi)\cdot(\phi/(1+\phi))=a\cdot((\phi\cdot\phi)/(1+\phi))=a\cdot1=a$. And since all 1-bits in the $(i+1)^\text{th}$ column comes after a 0-bit, they all naturally have the same length. This concludes the inductive argument. \qed

\begin{figure}[t]
\centering
\includegraphics[width=0.73\linewidth]{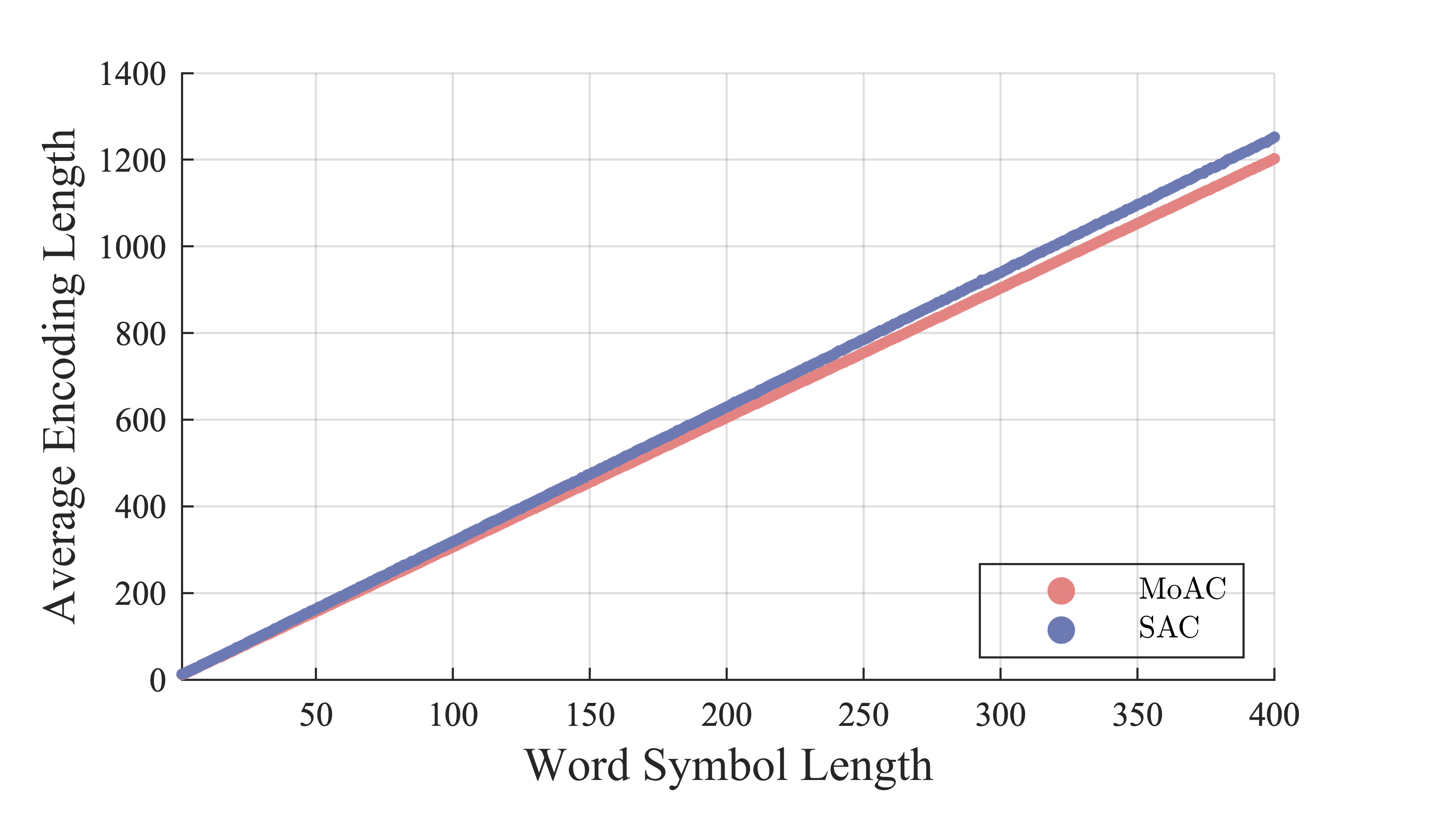}
\caption{Average Encoding Length of MoAC and SAC}

\vspace{-7pt} 
\end{figure}

\begin{figure}[t]
\centering
\includegraphics[width=0.73\linewidth]{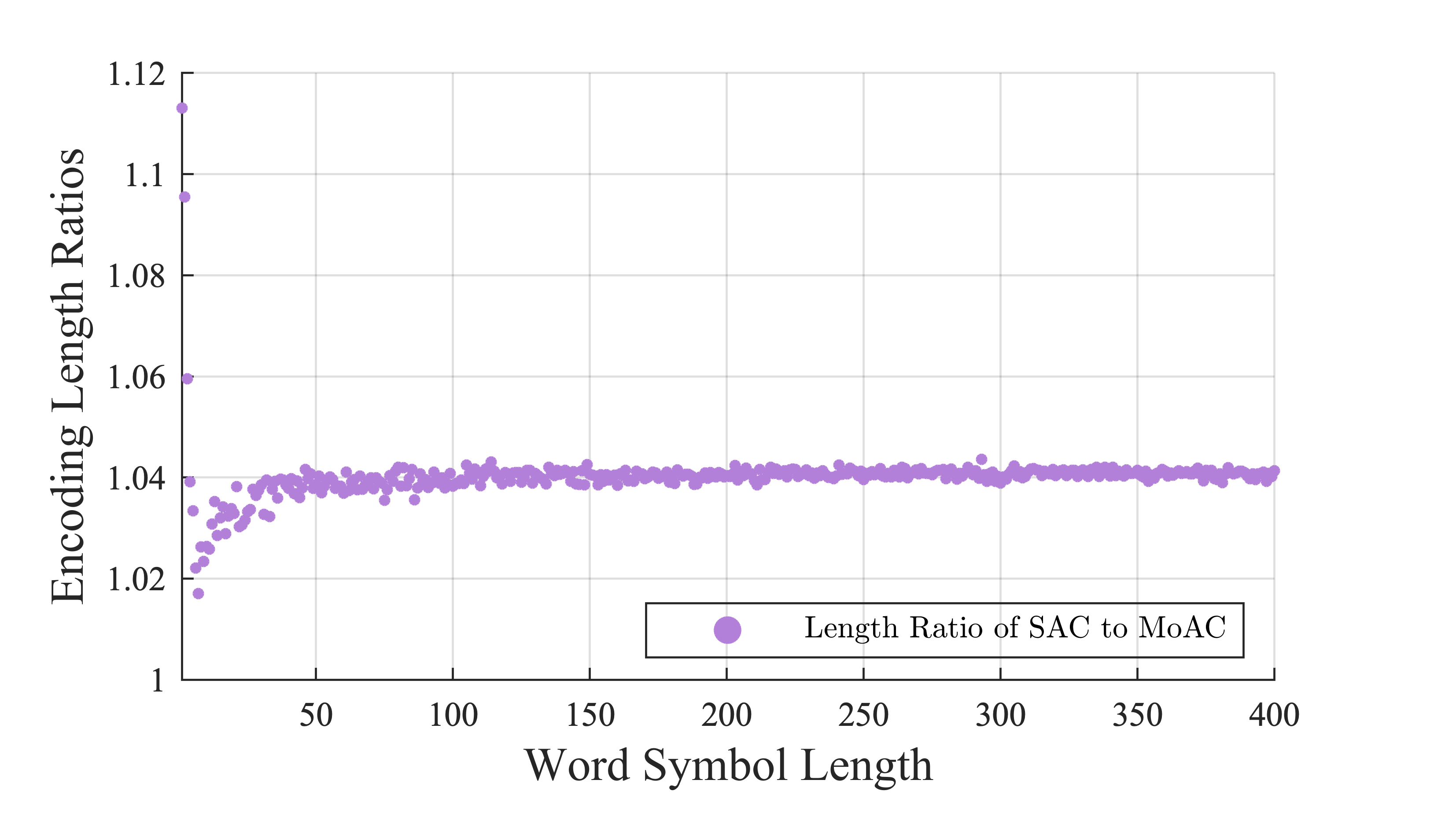}
\caption{The ratios of the Average Encoding Lengths of SAC to those of MoAC}

\vspace{-2pt}
\end{figure}
\vspace{-1pt}
Assume for any given $word$ of any alphabet, the interval $[a, b)$ is assigned to it using (4). Let $x$ be the height of the interval assigned to the $word$ (i.e., $x=b-a$). Then for an AC code to be assigned to $word$, at least, the interval height of this code must not be bigger than $x$. Thus the shortest AC  code that can be assigned to $word$ has the length \( \lceil \log_{\frac{1}{2}} x \rceil\). Since in an AC code, appearance of 1-bits and 0-bits are equally likely, on average, an AC code of length $n$ is transformed into a SAC code of length $(1/2)\cdot n+2\cdot(1/2)\cdot n = (3/2) \cdot n$. Thus, for the given $word$, its shortest expected SAC encoding length is $ \lfloor  (3/2)\cdot  \lceil \log_{\frac{1}{2}} x \rceil \rfloor$.

The MoAC encoding of $word$ must end with a 0-bit; and from the Lemma and (7), all the MoAC codes of length $n$ that ends with a 0-bit has an assigned interval height of $(1/\phi)^n$. Due to how MoAC code space is defined, if the equation $(1/\phi)^{n} \le x/2$  holds, then it guarantees that there exist a MoAC code of length $n$ whose interval $[k, l)$ is a subset of the interval $[a, b)$ associated with $word$. This equation then implies that the upper bound on the length of MoAC encoding of $word$ is \( \lceil (log_{\frac{1}{\phi}} x/2) \rceil \le \lceil (log_{\frac{1}{\phi}} x) \rceil +2\). In MoAC, a 0-bit is appended to an encoding that ends with a 1-bit. Thus the final upper bound becomes $\lceil (log_{\frac{1}{\phi}} x) \rceil +3$.

Note that, for any finite alphabet that consists of more than one symbol, the following fact can easily be proven: For any infinitesimally small positive real number \(\epsilon\), there exist a natural number $N$, such that the height values of the intervals, assigned to all words lengthier than $N$, are smaller than \(\epsilon\). Thus, for an alphabet containing more than one symbol, and for all of its sufficiently lengthy words, (8) gives the ratio of expected encoding length of SAC to that of MoAC:
\vspace{-2pt}
\vspace{2pt}
\begin{equation}
\lim_{{x \to 0^{+}}} \frac{\lfloor  (3/2)\cdot  \lceil \log_{\frac{1}{2}} x \rceil \rfloor}{\lceil (log_{\frac{1}{\phi}} x) \rceil+3} = (3/2)\cdot \log_{2} \phi \approx 1.0413...
\label{eq:golden_ratio}
\tag{8}
\vspace{1pt}
\end{equation}
\vspace{-6pt}

\vspace{-0.2cm}
To compare the average number of 1-bits produced by SAC and MoAC, we first calculate the appearance probability of 1-bits in a MoAC code by counting the 1-bits in $C(n)$. Let $one[n]$ denote the total number of 1-bits in $C(n)$. From (5), $one[1]=1$, and $one[2]=2$. We remarked that $C(n)$ can be obtained by inserting 0 to the start of all $C(n-1)$ and by inserting 10 to the start of all $C(n-2)$. Therefore we have $one[n]=one[n-1]+one[n-2]+|C(n-2)|=one[n-1]+one[n-2]+Fibonacci[n]$. The sequence $one[n-1]$ is known as the self-convolution of the Fibonacci numbers \cite{Fibonacci_article}.

\begin{figure}[t]
\centering
\includegraphics[width=0.73\linewidth]{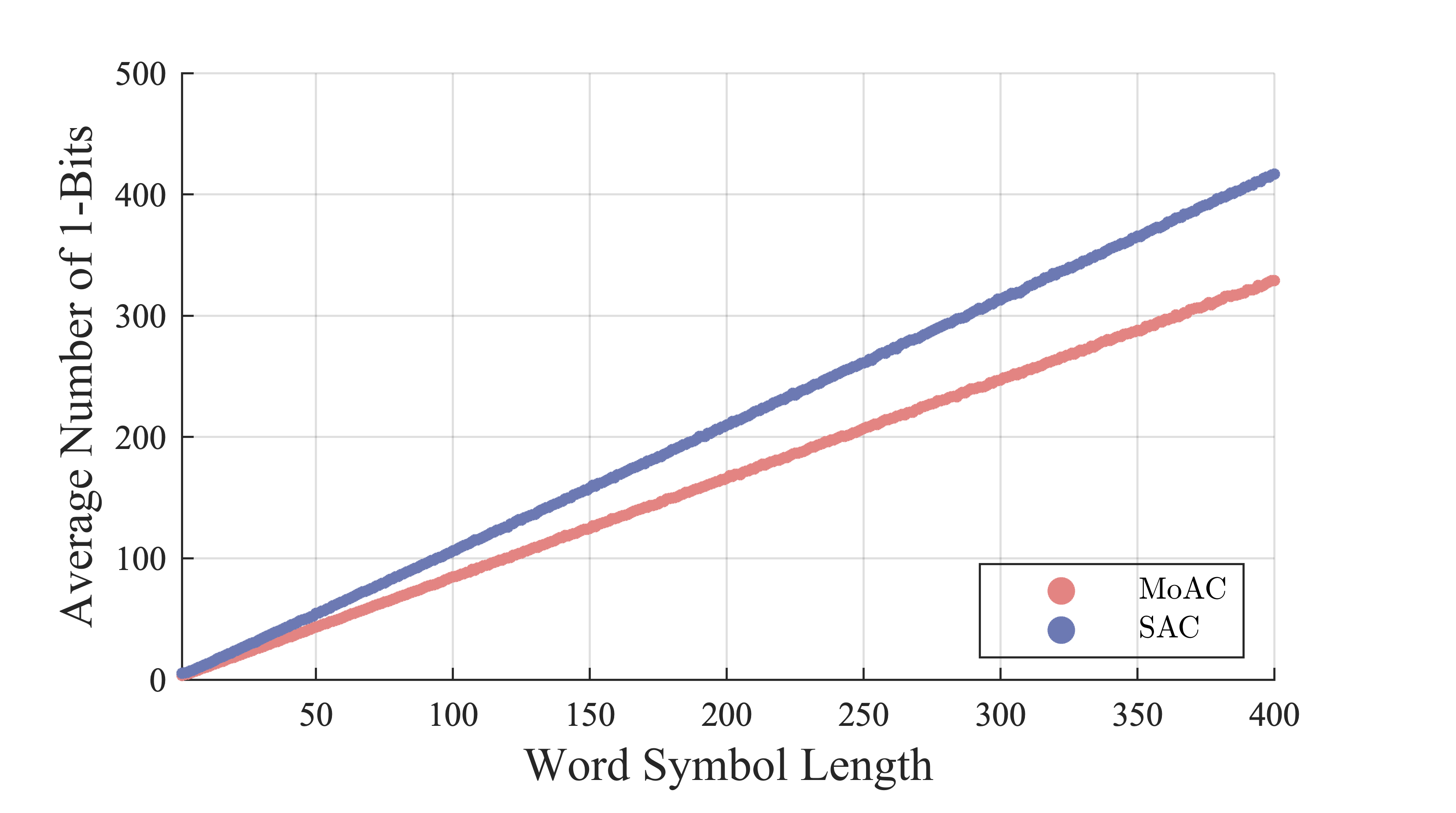}
\caption{Average Number of 1-bits of MoAC and SAC}

\vspace{-7pt} 
\end{figure}

\begin{figure}[t]
\centering
\includegraphics[width=0.73\linewidth]{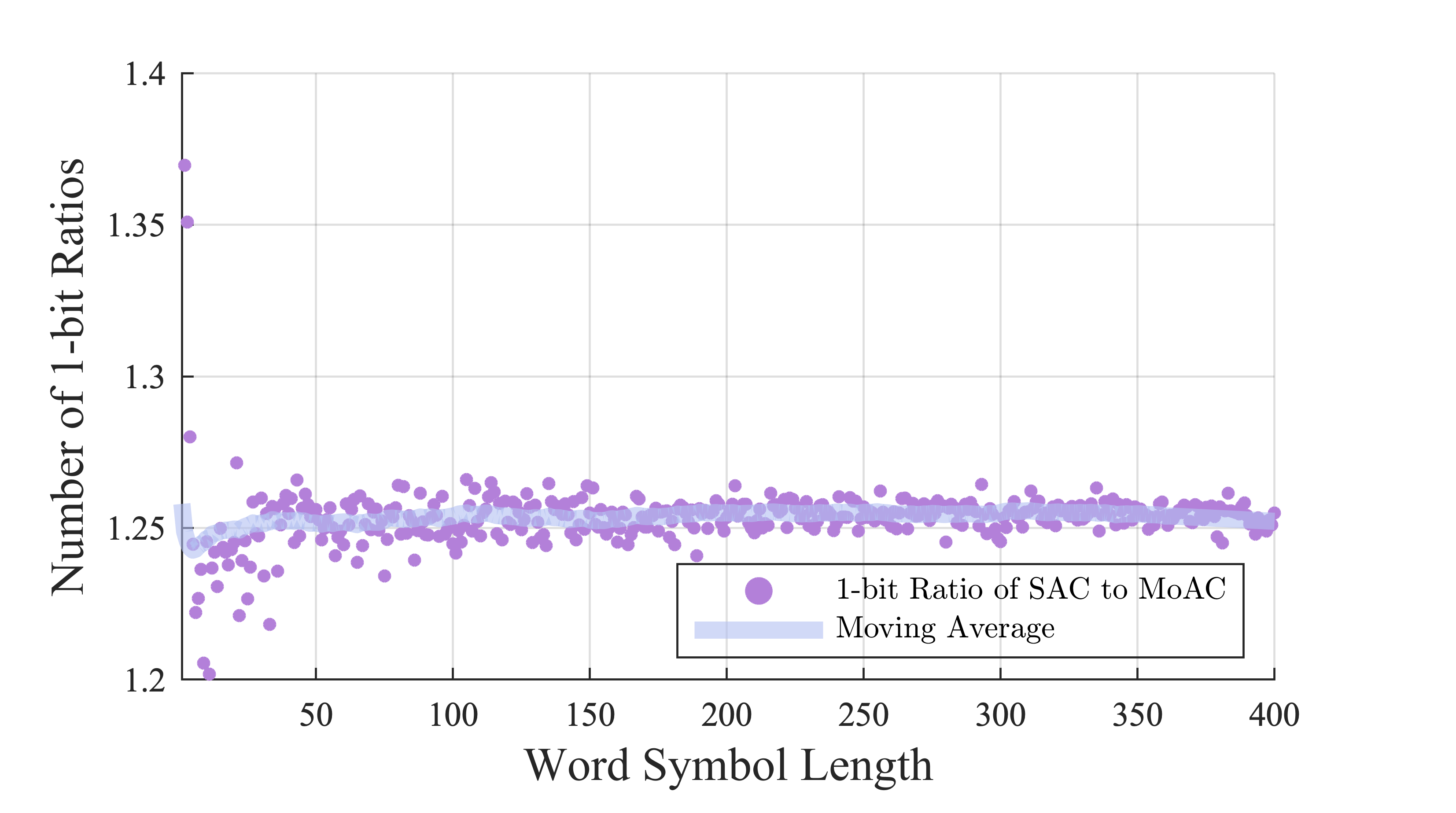}
\caption{The ratios of the Average Number of 1-bits of SAC to those of MoAC}

\vspace{-2pt}
\end{figure}

We remind that in a MoAC code of length \( n \), all codes must end with a 0-bit. Hence, for a MoAC code of length \( n \), all possible codes are the elements of \( C(n-1) \), each appended with a 0-bit. Thus, using the Lemma, all MoAC codes of length \( n \) appear with equal probabilities (i.e., they have the same interval heights). Consequently, \( \text{one}[n-1] \) gives the expected number of all 1-bits in all MoAC codes of length \( n \). Note that the total number of bits in all MoAC codes of length \( n \) is given by \( n \cdot |C(n-1)| = n \cdot \text{$Fibonacci$}[n+1] \). Accordingly, the following limit, $l$, which we have computed using numerical methods, gives the expected ratio of 1-bits in a MoAC code:

\vspace{1pt}
\vspace{-0.1cm}
\begin{equation}
l=\lim_{{n \to \infty}} (\text{$one$}[n-1] / (n \cdot \text{$Fibonacci$}[n+1])) \approx 0.276...
\label{eq:golden_ratio}
\tag{9}
\vspace{-0.1cm}
\end{equation}
\vspace{-12pt}
\vspace{1pt}

\vspace{-0.00cm}
Recall that in SAC, each 1-bit produced by AC is replaced with a 10. Since in AC, the distribution of 1-bits and 0-bits are equally likely, the appearance probability of 1-bit in SAC is 1/3, while that of a 0-bit is 2/3.  The expected number of 1-bits in MoAC for a word, of which encoding length is $n$, is $l\cdot n \approx 0.276 \cdot n$. For the same word, the number of expected 1-bits in its SAC encoding, from (8), is $(3/2)\cdot (\log_{2} \phi) \cdot n\cdot (1/3)$. Dividing these two numbers we get $((3/2)\cdot (\log_{2} \phi) \cdot n\cdot (1/3))/(0.276 \cdot n)$ $\approx 1.257$. This shows that SAC (and AC), in average, uses $25.7$ percent more 1-bits than MoAC does.

To reinforce these theoretical results, finite precision versions of MoAC and SAC will now be compared. Let our exemplary alphabet be $(A, B, C, EOF)$, with corresponding respective probabilities $(0.33, 0.33, 0.33, 0.01)$. The average encoding length and number of 1-bits comparisons are given in Figs. 7 and 9 respectively. For each word length, we chose 400 random words using the symbol distributions of the exemplary alphabet. And a bit precision of 20 was designated. Fig. 8 and 10 empirically verify the theoretical length and 1-bit ratios of $1.0413$, and $1.257$ respectively. 
\vspace{1pt}
\subsubsection{Finite Precision Zero-Order MoAC}
Due to the unsymmetrical nature of MoAC as can be seen in Fig. \ref{fig3}, the implementation of finite precision MoAC is non-arbitrarily different, and more complex than the finite-precision implementation of AC given in \cite{arithmetic_coding}, \cite{data_compression_book}. The link for a GitHub repository that includes the zero-order Python implementations and pseudo-codes of MoAC and AC (both with and without EOF versions) are provided in the Code Availability section.

\vspace{1pt}
\subsubsection{Finite Precision Higher-Order MoAC}
Once an algorithm for the zero-order MoAC is available, implementation of the higher order MoAC is trivial. We just allocate intervals for symbols according to their conditional probabilities, based on the preceding symbols. Corresponding changes in MoAC decoder can also be easily implemented. Other than this, there is no need for change in any other part of the proposed MoAC algorithm. For a better understanding, interested readers may look into the Higher-Order Modeling chapter of the book \cite{data_compression_book}.
\vspace{-0.6cm}
\vspace{-2pt}
\subsection{Molecular Arithmetic with Prefix Coding (MoAPC)}
\label{sec:MoAPC}
\vspace{-0.1cm}
\vspace{2pt}

Since the number of precision bits has to be finite, unique decodability of MoAC is not guaranteed. In the Python implementation of MoAC, we have created an underflow expansion process which is non-arbitrarily different than that of the AC. Although this measure increases the accurate decodability rate of MoAC, there can still be cases where decoding errors do occur. Hence, the encoder is required to decode the MoAC encoding of the to-be-transmitted word to verify a perfect matching between the original and the resultant words. The implementation of this checking mechanism can be found in the GitHub repository, whose link is provided in the Code Availability section.  If resultant words do not match, the word is encoded through MoPC$^{*}$. We name this as the Molecular Arithmetic with Molecular Prefix Coding (MoAPC).

So, there needs to be a mechanism to inform the decoder if the incoming message was encoded through MoAC or MoPC$^{*}$. For this purpose, the header mechanism shown in Fig. 11 is proposed to inform the decoder which encoding scheme was opted for. For MoAC, encoding a word, and decoding its encoding have almost the same computational cost. However, since the decoder just repeats the steps of the encoder in our implementation of MoAC, if the transmitter has a strong memory component, the decoding inside it can be computationally more efficient. In our proposed header mechanism, a 0-bit is inserted to the start of a word if it was encoded by MoAC, and the bit sequence 10 is inserted to the start of a word if it was encoded by MoPC$^{*}$.
\vspace{3pt}
\begin{figure}[H]
\centering
\includegraphics[width=0.77\linewidth]{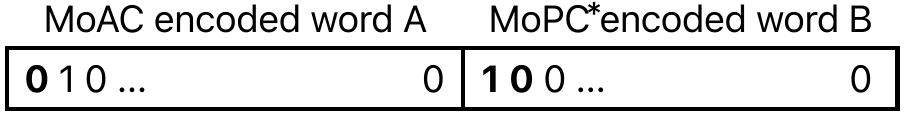}
\caption{ MoAPC: MoAC / MoPC$^{*}$ Distinguisher}

\end{figure}
\vspace{-3pt}

\vspace{-0.2cm}
\section{Detection}
\label{detection}

\subsection{Algorithms for Detection and Error Correction}
\label{sec:detection}

We follow an almost-identical detection approach to the one introduced in \cite{best_channel_coding_2020}. Please note that the only original contributions in this subsection are the introduction of the $min$ constant, redefinition of $r_i^{min}$ \footnote{For ISI-Mitigating Codes \cite{best_channel_coding_2020}, the parameter $r^i_{min}$ is defined as $min(\mathbf{r}^{\bm{i}}\setminus\{r_1^i\})$. By redefining $r^i_{min}$ as being $non\_zero\_min(\mathbf{r}^{\bm{i}})$, we have generalized its detection mechanism to be applicable across various coding schemes. In the Performance Evaluation section, to ensure a fair comparison, we tested ISI-Mitigating Codes \cite{best_channel_coding_2020} using both the redefined and original approaches. As shown in Figs. 21–24, both methods yield comparable error rates.}, the definition of the last chunk of the bit-sequence, $\mathbf{b}^{\lfloor \bm{n}/\bm{spacing} \rfloor}$, and the optimization of the $spacing$ constant, that was previously assigned a fixed value based on the coding block length used.

Let $\mathbf{r}^{\bm{i}}=(r_{1}^i,r_{2}^i,...,r_{spacing}^i)$ represent the count of the detected information molecules for corresponding signal intervals for the incoming message $ \mathbf{b}^{\bm{i}}= (b_{spacing\cdot(i-1)+1},...,b_{spacing\cdot(i-1)+spacing})$, where $b_{j}$ denotes the $j^\text{th}$ bit of the whole encoded message, and $spacing$ is an integer constant.  If the number of bits of the whole encoding is $n$, and if $spacing$ does not divide $n$,  $\mathbf{b}^{\lfloor \bm{n}/\bm{spacing} \rfloor}$ is defined to be the last $(spacing$ $+$ $n$ $mod$ $spacing)$ bits of the whole encoding. Then, we can similarly define $\mathbf{r}^{\lfloor \bm{n}/\bm{spacing} \rfloor}$. The integer constant $spacing$ can take the value that results in the least symbol error rate value.
 
\setlength{\abovedisplayskip}{12pt}  
\setlength{\belowdisplayskip}{5pt}
 
 Define $r^i_{max} = max(\mathbf{r}^{\bm{i}})$, and $r^i_{min} = non\_zero\_min(\mathbf{r}^{\bm{i}})$. If $\mathbf{r}^{\bm{i}}=(0,0,...,0)$, take $r^i_{min}$ to be $\infty$. Then the optimal threshold, $\tau ^{i}$, of the $i^\text{th}$ code-word, $\mathbf{b}^{\bm{i}}$, can be found as \vspace{-0.3cm} \begin{equation*} \tau^{i} = a\cdot r^i_{min} + (1-a)\cdot r^i_{max} \tag{10}, \end{equation*}   where $a$ is the scaling coefficient \cite{best_channel_coding_2020}. Note that $0\leq a \leq1$. Most importantly, this scheme assumes that each  $\mathbf{b}^{\bm{i}}$ contains at least one 1-bit. But in source coding this may not always be the case. We solve this problem by introducing another channel-specific constant $min$ which denotes the least number of molecules that a receiver could detect in the signal interval of a 1-bit. In the proposed Algorithm 1, If the number of molecules in a signal-interval falls below $min$, that signal interval is always detected to be a 0-bit.

For determining $a$, we adopt the pilot-signal approach given in \cite{best_channel_coding_2020}, where, at the start of the communication, predetermined ensuing words are sent. Then, starting from 0 and continuing to 1, with a step size of $0.004$, the decoder can determine the value of $a$ that results in the most accurate decoding of predetermined pilot symbols in terms of symbol error rate. After the incoming message is detected using the threshold method given in Algorithm 1, detected bit sequence is processed through an ISI-mitigating error correction algorithm defined in \cite{best_channel_coding_2020}, and given in Algorithm 2. This algorithm, taking the advantage of the fact that the proposed coding does not {\parfillskip=0pt \par}

\begin{algorithm}[H]
     
\caption{ Detection Algorithm }
\begin{algorithmic}[1]
\Require the $molecule\_count\_sequence$ with size $n$, where $molecule\_count\_sequence$[i] denotes the number of molecules detected at the $i^\text{th}$ signal interval, the scaling coefficient $a$, the spacing constant $spacing$, and the minimum constant $min$
\State let $detected\_bit\_sequence$ be a sequence of $0s$ of size $n$

\For{$k \gets 1$ \textbf{to} $n$}

    \State $i=\lfloor (k-1))/spacing \rfloor+1$
    \If{$molecule\_count\_sequence[k]\ge \tau^i$} 
    \If{$molecule\_count\_sequence[k]\ge min$}
    \State $detected\_bit\_sequence[k]=1$
    \EndIf
    \EndIf
\EndFor

\State \Return $detected\_bit\_sequence$ 
\end{algorithmic}

\end{algorithm}

\vspace{0.1cm}

\begin{algorithm}[H]

\caption{ Error Correction Algorithm \cite{best_channel_coding_2020} }
\begin{algorithmic}[1]
\Require  $detected\_bit\_sequence$ with size $n$

\For{$j \gets 1$ \textbf{to} $n$}
    \If{$detected\_bit\_sequence[j]==1$ and  \newline
    \hspace*{1.5em} not $(j==n)$}
    \State $detected\_bit\_sequence[j+1]=0$
    \EndIf
    \EndFor
\end{algorithmic}
\end{algorithm}

\hspace{-13pt} contain consecutive 1-bits, mitigates the ISI that may be caused by a preceding 1-bit. 

\vspace{-0.5cm}
\vspace{-3pt}
\vspace{0.1cm}

\vspace{9pt}
\subsection{Word Differentiation for EOF-Excluded Words}
\vspace{-2pt} 
If two types of information molecules are available at the transmitter, words belonging to an EOF-excluding alphabet can be distinctively transmitted, irrespective of the encoding scheme used: To differentiate ensuing words, a 1-bit is transmitted at the beginning of each word's transmission using type-2 (coloured in blue) molecules, while type-1 (coloured in red) molecules are exclusively used for transmitting the encoding of each word, as shown in Fig. 12.

\vspace{4pt}

\begin{figure}[H]
\centering
\includegraphics[width=0.9\linewidth]{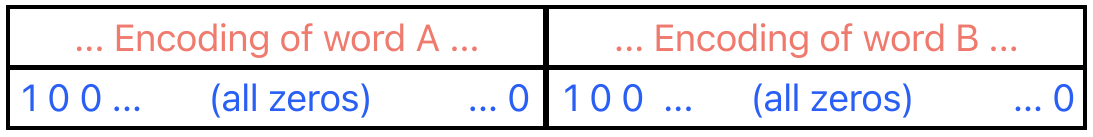}
\caption{2 Molecule Types Word Distinguisher}
\vspace{-0.1cm}
\end{figure}

\vspace{-0.1cm}
\section{Performance Evaluation}
\label{sec:third}
\subsection{Encoding Length and Power Consumption Comparisons}
\vspace{-0.05cm}
In Alphabets 1 and 2, we represent exemplary nucleotide probability distributions of a single-strand DNA. For the Alphabet 1, we have chosen the probability values to create a non-uniform (i.e., a lower entropy) symbol distribution. In contrast, for Alphabet 2, we have selected an alphabet with a more uniform distribution (i.e., a higher entropy). This selection allows us to more thoroughly asses the performance of our proposed methods as the performance of coding schemes can vary between nearly-uniform and non-uniform symbol alphabets \cite{data_compression_book}. Alphabet 1 does not contain an EOF symbol while Alphabet 2 contains an EOF symbol. Testing the performance of MoAC (thus that of MoAPC) for both EOF-included and EOF-excluded cases are important, as the finite-precision implementation of MoAC, which is accessible in the Code Availability section, is different between EOF-included and EOF-excluded versions.

In order to minimize the expected power consumption of  ISI-Mitigating, Uncoded, and Huffman coding methods, we have assigned codes that contain the fewest number of 1-bits to the symbols with the highest probabilities. For each compared method, average encoding length and average number of 1-bits comparisons  for words of length from 1 to 400 are given in Figs. 13, 14, 17, and 18. For each word length, we  randomly chose 400 words using the symbol distributions of the corresponding alphabet. For MoAC, AC and SAC, bits precision of 20 is designated. In both alphabets, availability of a single type of an information molecule is assumed; and the mechanism shown in Fig. 11 is adopted for MoAPC.

In Figs. 15 and 19, we compare the average encoding lengths of all error-correcting compression methods to the average encoding length of MoAPC. As the figures show, MoAPC has a shorter average encoding length than all these methods. During additional comparisons with other symbol alphabets (which are not shown here for the sake of brevity), MoAPC consistently had a shorter average encoding length compared to MoPC$^{*}$, MoHuffman \cite{lee2023isimitigating}, SAC, and ISI-Mitigating codes \cite{best_channel_coding_2020} for words longer than an alphabet-dependent number, which is usually less than $50$. In Fig. 19, MoPC$^{*}$ is better than MoHuffman for word lengths less than $81$; however, for word lengths greater than $81$, it is mostly outperformed by MoHuffman. The reason for this is that the probability of the EOF symbol for Alphabet 2 is set at $0.05$. As the word length increases beyond $20$, the probability of the EOF symbol decreases, leading to a change in the actual probability distribution of the alphabet. This causes the MoPC$^{*}$ to perform in an alphabet distribution for which it was not optimized, leading to a slight reduction in its performance.


For each alphabet, in Figs. 16 and 20, accurate decoding ratio of  arithmetic coding methods are given. As can be observed in these figures, each encoding that AC (and thus SAC) produces, can almost always be correctly decoded. However this is not the case for MoAC, whose accurate decoding ratio decreases as the word symbol length increases, due to its irrational nature. This phenomena further justifies the use of the uniquely decodable MoAPC. In terms of the power consumption, for the Alphabet 1, MoAPC outperforms all given methods, including its arithmetic coding competitors AC and SAC, except the Uncoded one in Fig. 14. In Alphabet 2, in addition to the Uncoded one, MoAPC is also outperformed by MoPC$^{*}$, as shown in Fig. 18. This power performance variance between MoPC$^{*}$ and MoAPC on the exemplary alphabets indicates the alphabet-dependent nature of source coding.

\vspace{-0.3cm}
\subsection{MC Channel Simulation Results} 
\label{sec:simulations}
\vspace{3pt}
\vspace{-0.1cm}

\begin{table}[t]
\caption{Exemplary Alphabet 1}
\centering
{\footnotesize
\begin{tabular}{c|cccc}
\hline
Symbol & A & T & C & G  \\
\hline
Probability & 0.50 & 0.25 & 0.23 & 0.02  \\
\hline
Uncoded & 00 & 01 & 10 & 11  \\
\hline
ISI-Mitigating \cite{best_channel_coding_2020} & 0001 & 0010 & 0100 & 0101  \\
\hline
Huffman & 0 & 10 & 110 & 111  \\
\hline
MoHuffman \cite{lee2023isimitigating} & 0 & 100 & 10100 & 101010  \\
\hline
MoPC$^{*}$ & 0 & 100 & 10100 & 101010  \\
\hline

\end{tabular}
}
\end{table}

\begin{figure}[t]
\vspace{-0.1cm}
\centering

\includegraphics[width=0.75\linewidth]{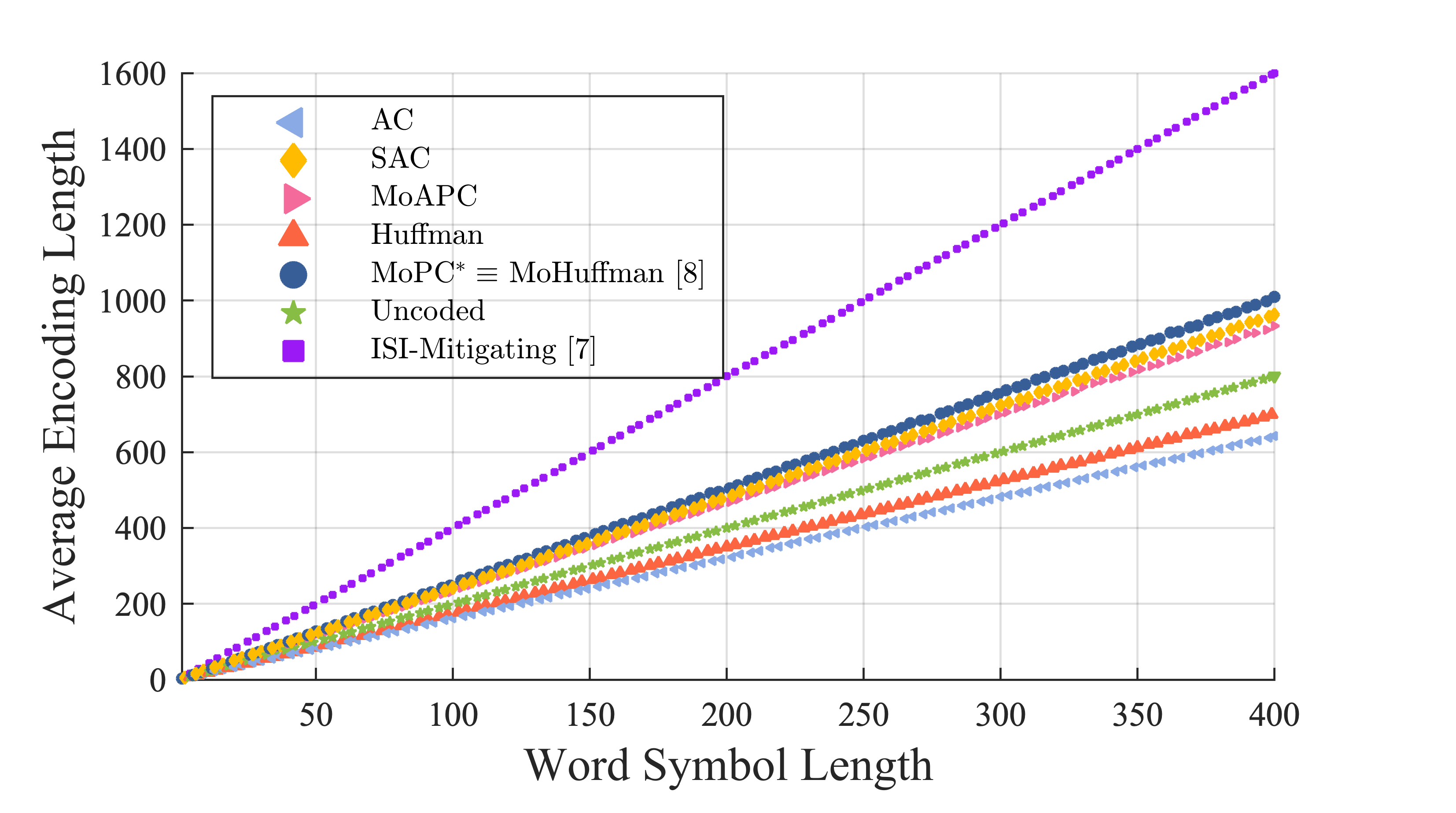}
\caption{Encoding Length Comparison for Alphabet 1}

\includegraphics[width=0.75\linewidth]{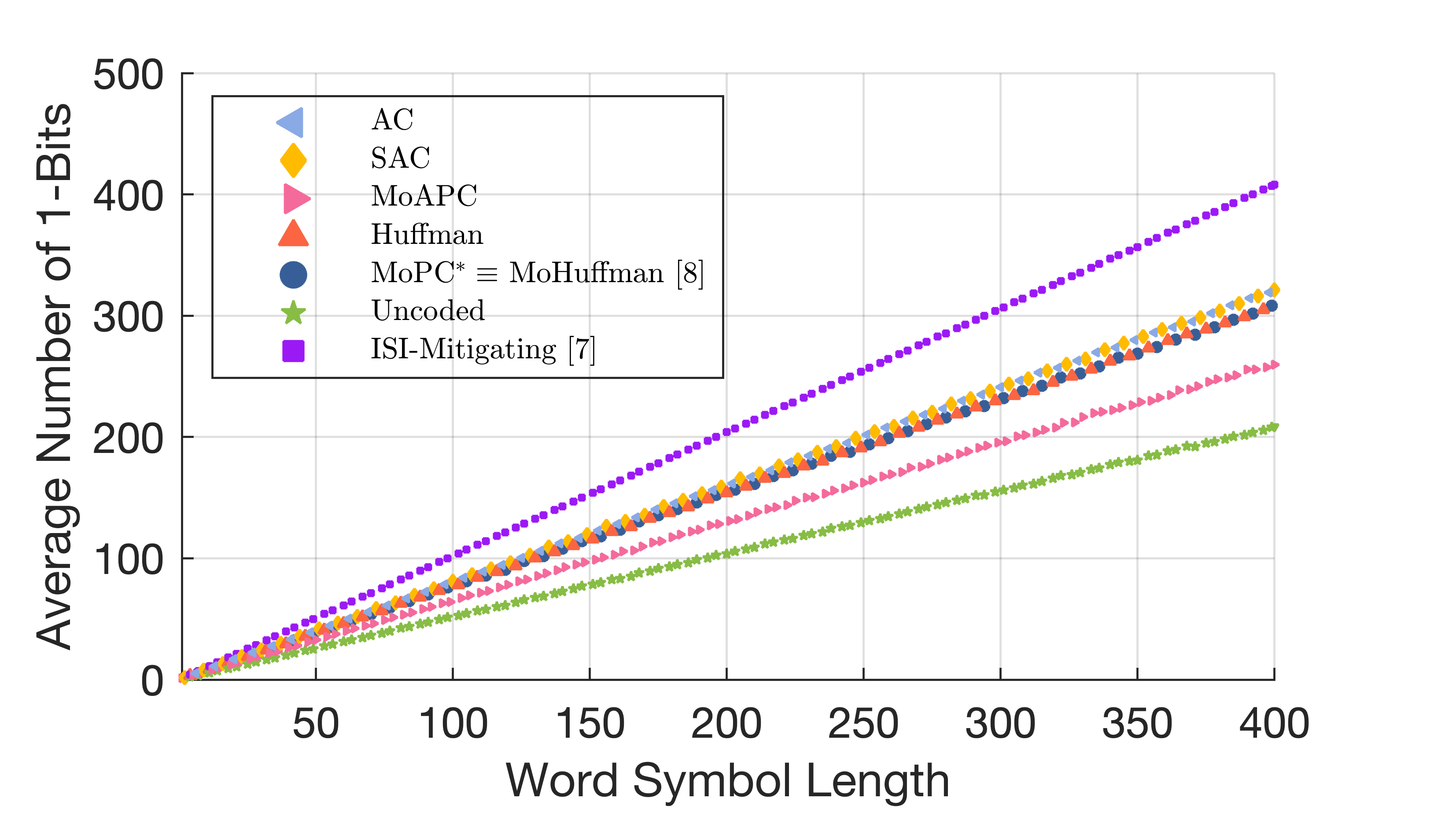}
\caption{Power Consumption Comparison for Alphabet 1}

\includegraphics[width=0.75\linewidth]{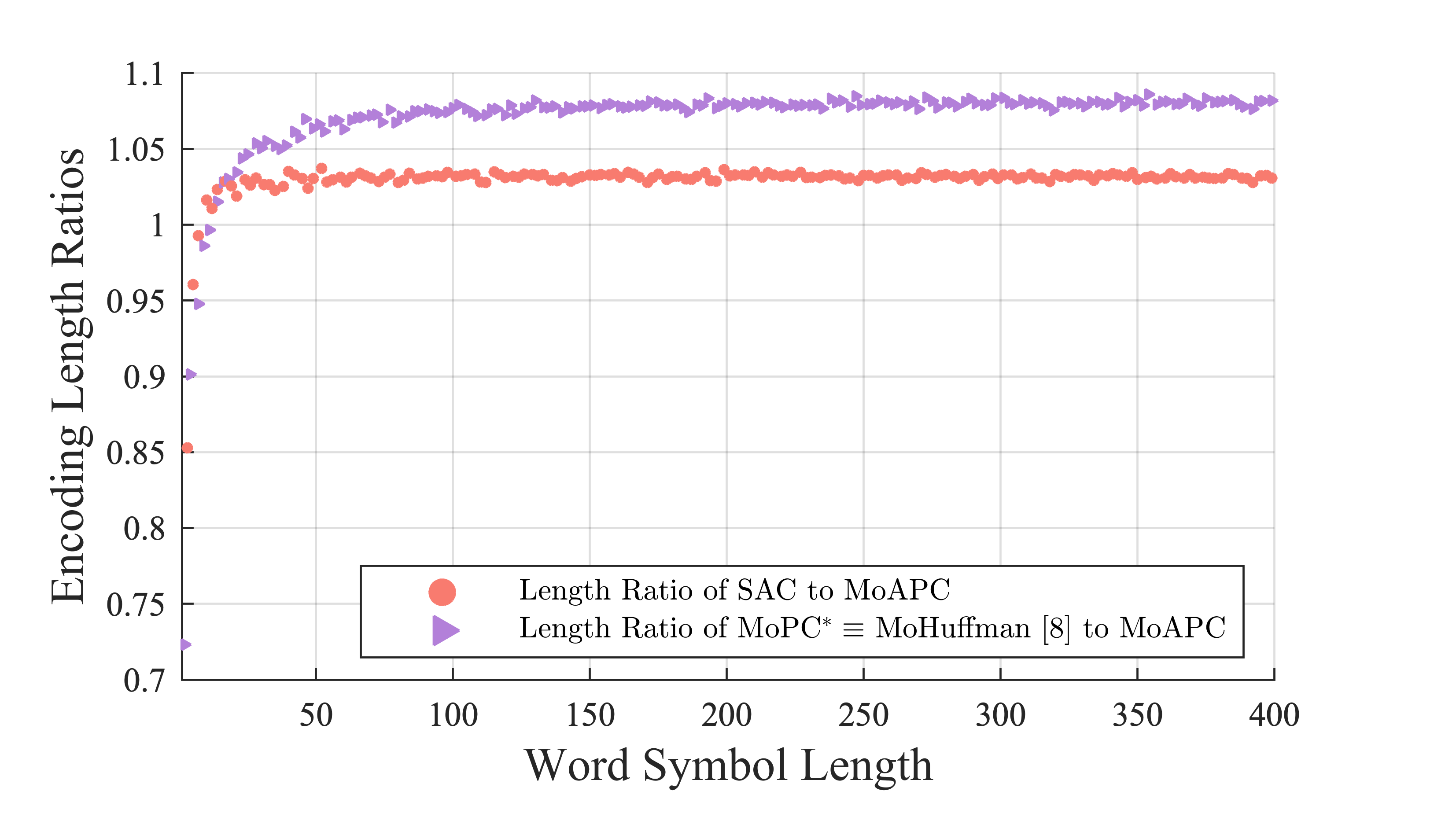}
\caption{Encoding Length Ratios for Alphabet 1}

\includegraphics[width=0.75\linewidth]{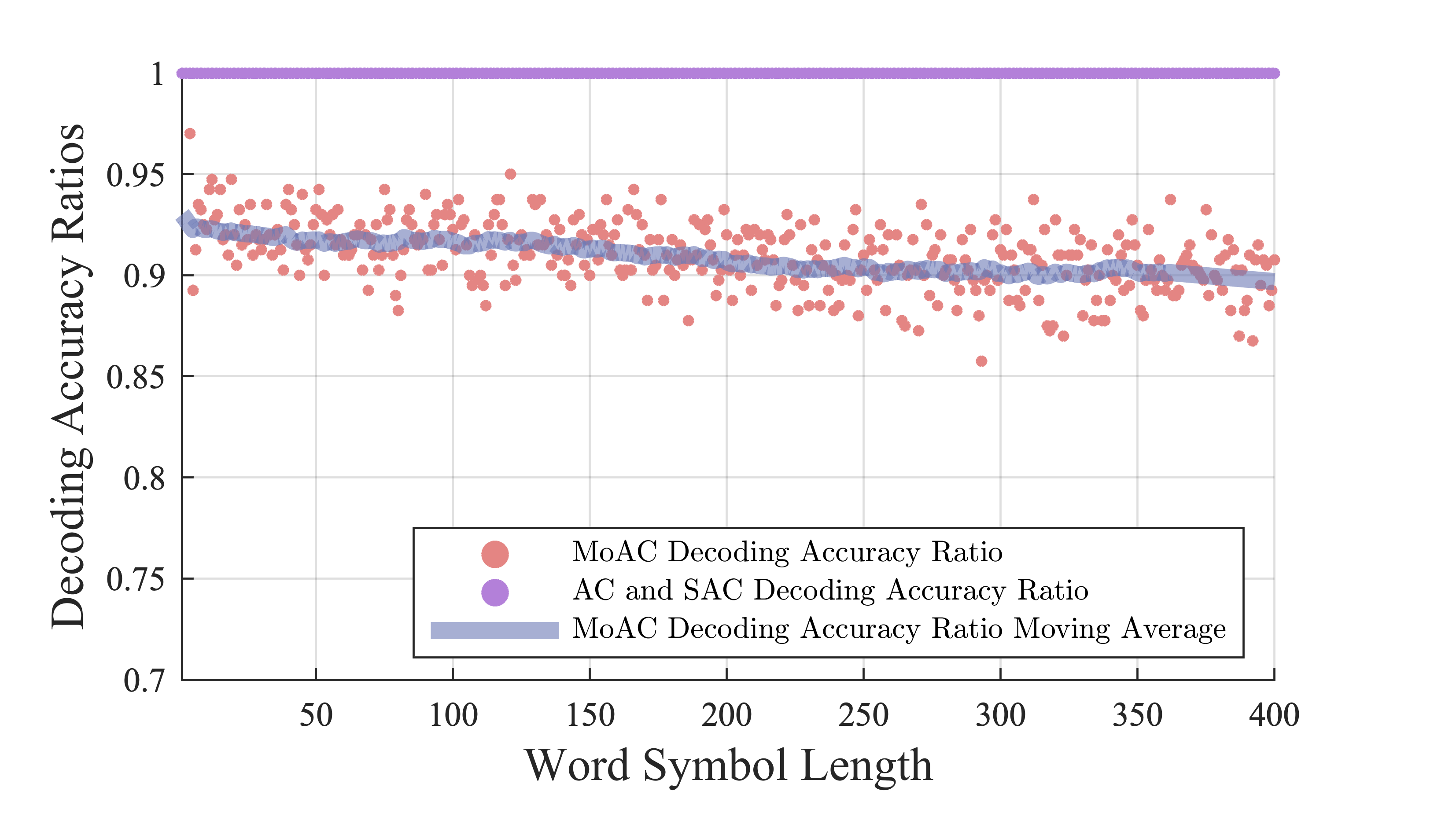}
\caption{Arithmetic Accuracy Ratios for Alphabet 1}

\end{figure}

\begin{table}[t]
\caption{Exemplary Alphabet 2}
\centering
{\footnotesize
\begin{tabular}{c|ccccc}
\hline
Symbol & A & T & C & G & EOF \\
\hline
Probability & 0.25 & 0.24 & 0.23 & 0.23 & 0.05 \\
\hline
Uncoded & 000 & 001 & 010 & 100 & 011 \\
\hline
ISI-Mitigating \cite{best_channel_coding_2020} & 00001 & 00010 & 00100 & 01000 & 00101 \\
\hline
Huffman & 00 & 10 & 01 & 110 & 111 \\
\hline
MoHuffman \cite{lee2023isimitigating} & 00 & 100 & 010 & 10100 & 101010 \\
\hline
MoPC$^{*}$ & 000 & 100 & 010 & 0010 & 1010 \\
\hline

\end{tabular}
}
\end{table}

\begin{figure}[t]
\vspace{-0.1cm}
\centering

\includegraphics[width=0.75\linewidth]{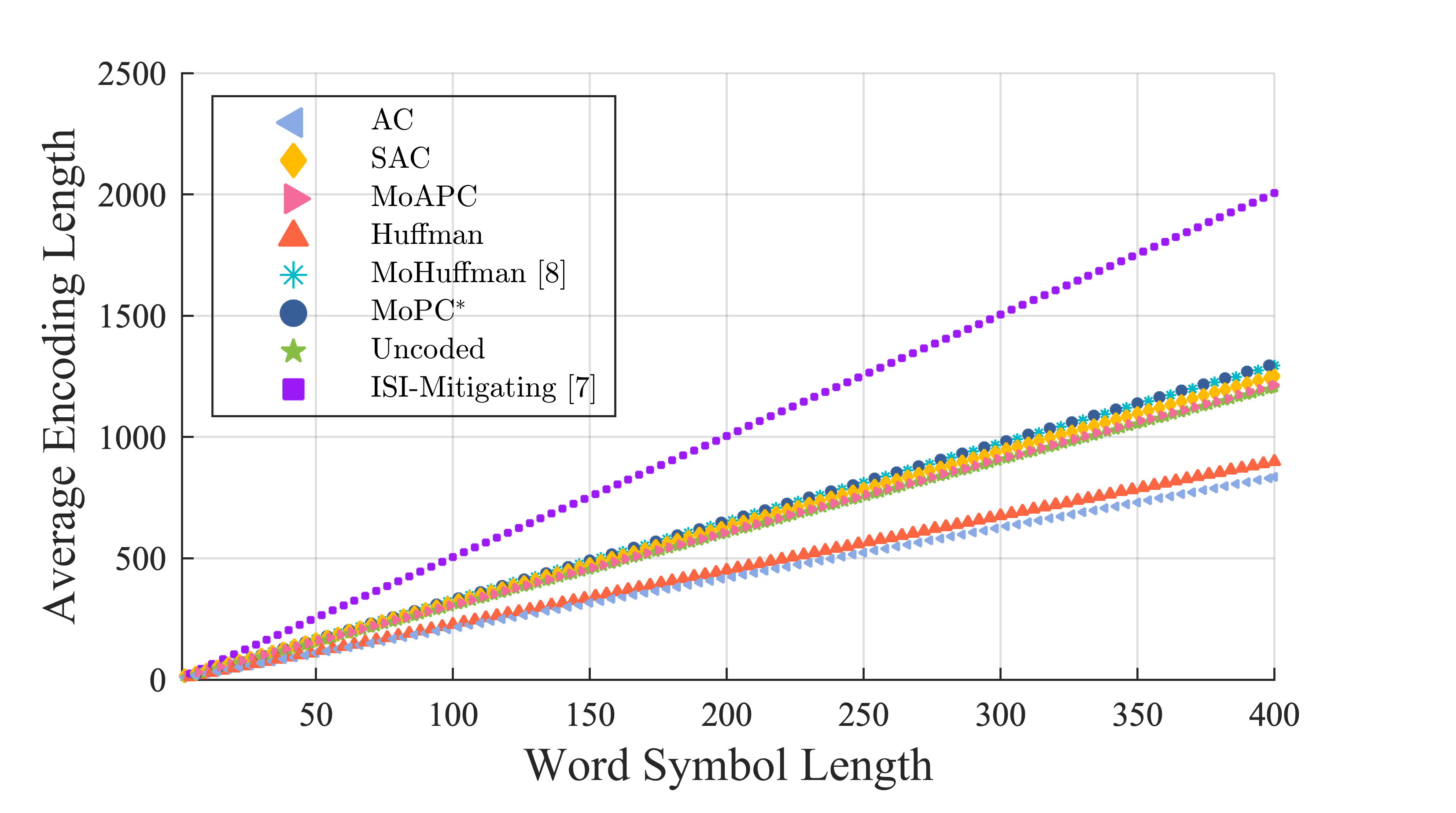}
\caption{Encoding Length Comparison for Alphabet 2}

\includegraphics[width=0.75\linewidth]{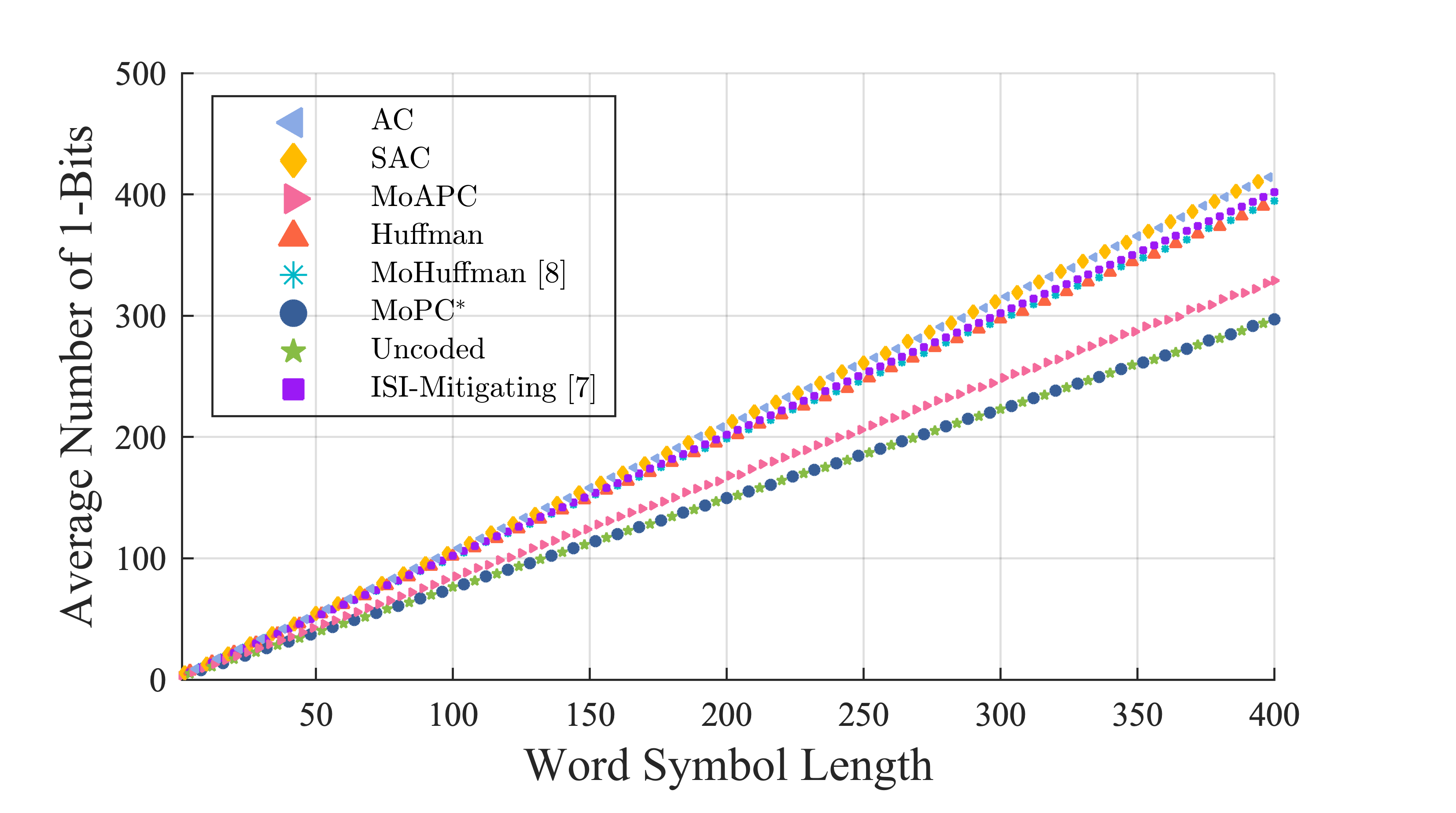}
\caption{Power Consumption Comparison for Alphabet 2}

\includegraphics[width=0.75\linewidth]{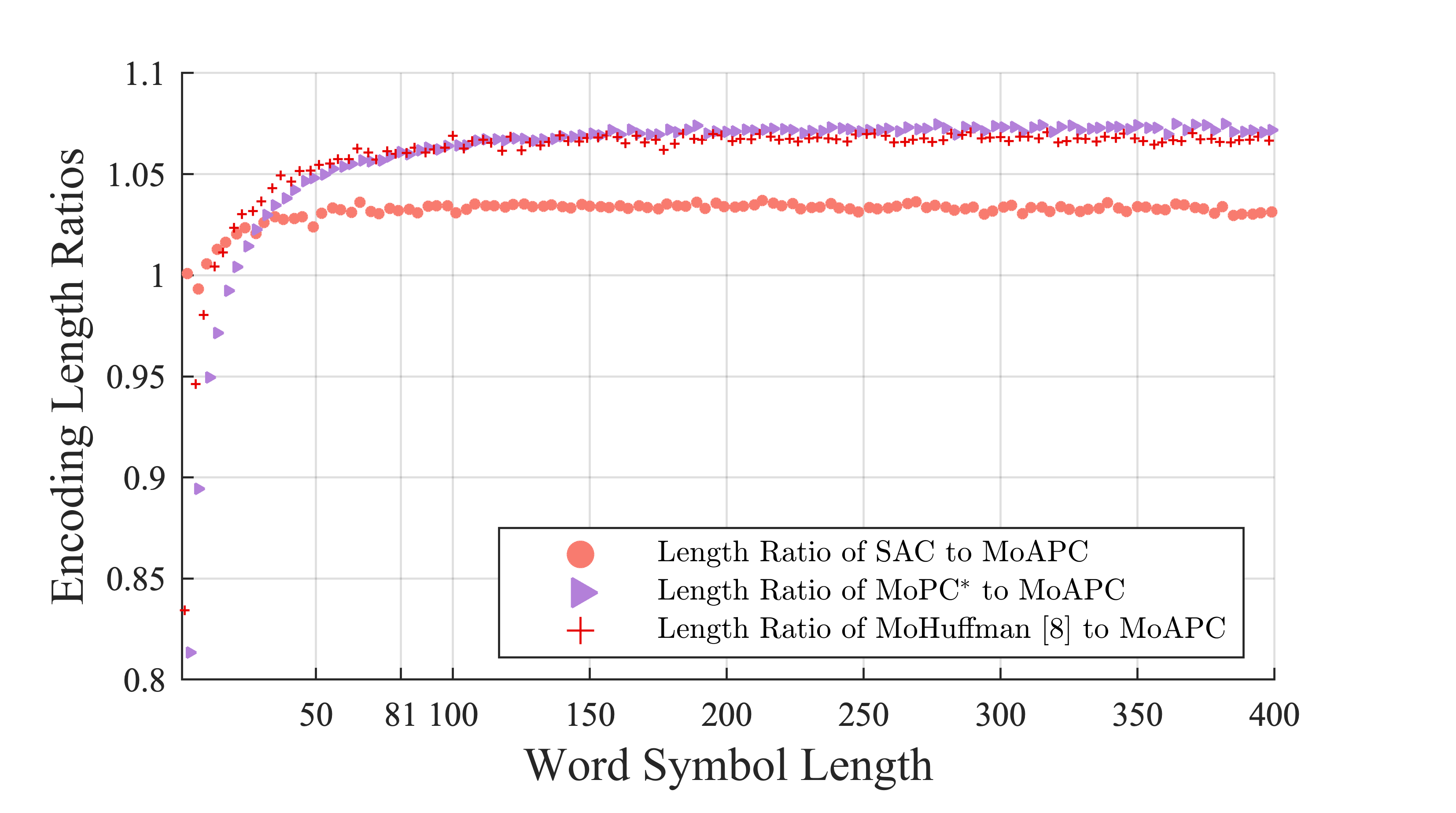}
\caption{Encoding Length Ratios for Alphabet 2}

\includegraphics[width=0.75\linewidth]{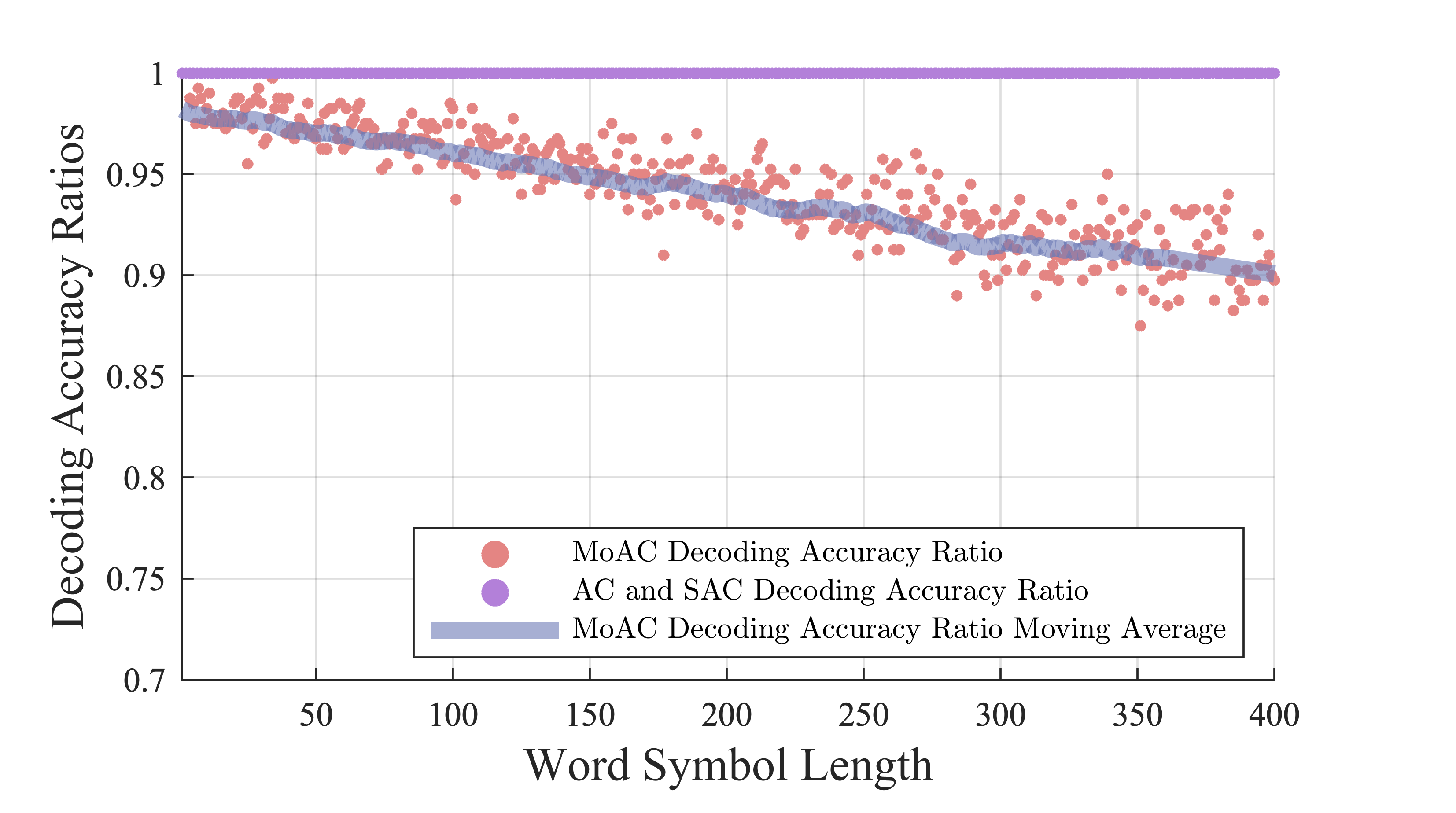}
\caption{Arithmetic Accuracy Ratios for Alphabet 2}

\end{figure}

In comparing different coding strategies for MC, normalizing the signal durations and signal powers is essential. This ensures that equal amounts of information are transmitted through various coding schemes within the same time duration while using an equal number of information molecules. The normalization is done in the following way, as briefly outlined in \cite{normalization}: Let $I$ be the set of all information (i.e., the words) available for transmission. In a deterministic approach, each element of $I$ appears with a probability of $1/|I|$. In a probabilistic approach, each element of \( I \) may appear with different probabilities, which sum to 1. Assume a coding scheme $C_1$ encodes a randomly chosen element of $I$, using $S_1$ expected number of bits, and $M_1$ expected number of 1-bits. Also assume that a coding scheme $C_2$ encodes a randomly chosen element of $I$, using $S_2$ expected number of bits, and $M_2$ expected number of 1-bits. Then the signal interval value of the coding scheme $C_2$ should be $S_1/S_2$ times the signal interval value of coding scheme $C_1$. Similarly, the molecule count per transmission of a 1-bit value for the coding scheme $C_2$ should be $M_1/M_2$ times that of the coding scheme $C_1$.

\begin{table}[t]
	\begin{center}
	\caption{Average Encoding Length and Power Consumption for Alphabet 1 with Word Length 20}
	\renewcommand{\arraystretch}{1.1}
	\label{tbl_system_parameters}
	\begin{tabular}{p{3.1cm}@{\hskip 0.2cm}c@{\hskip 0.5cm}c}
	\hline
	\bfseries{Coding Method} & {\centering \bfseries{Encoding Length}} & \parbox[c]{2.5cm}{\centering \bfseries{Power Consumption} \\ \footnotesize (Number of 1-bits)} \\ 
	\hline
	Uncoded & $$40$$ & $$10.4$$ \\
 ISI-Mitigating \cite{best_channel_coding_2020} & $$80$$ & $$20.4$$ \\
 AC  & $$33.32009$$ & $$16.53598$$ \\
 SAC & $$49.85607$$ & $$16.53598$$ \\
 MoAPC  & $$48.63674$$ & $$13.44275$$ \\
 Huffman  & $$35$$ & $$15.4$$ \\
	 MoPC$^{*}$ $\equiv$ MoHuffman \cite{lee2023isimitigating}  & $$50.4$$ & $$15.4$$ \\

	\hline
	\end{tabular} 
	\end{center}
	\renewcommand{\arraystretch}{1}
\vspace{15pt}
\end{table}

\begin{table}[t]
	\begin{center}
	\caption{Signal Interval and Molecule Count Normalizations for Alphabet 1 with Word Length 20}
	\renewcommand{\arraystretch}{1.1}
	\label{tbl_system_parameters}
	\begin{tabular}{p{3.1cm}@{\hskip 0.2cm}c@{\hskip 0.5cm}c}
	\hline
	\bfseries{Coding Method} & \parbox[c]{2cm}{\centering \bfseries{Signal Interval}} & \parbox[c]{2.5cm}{\centering \bfseries{Molecule Count} \\ \footnotesize ($M=100, 200,...$)} \\ 
	\hline
    Uncoded & $200\, $ms & $1\cdot M$ \\
    ISI-Mitigating \cite{best_channel_coding_2020} & $100\, $ms & $\left\lfloor 0.5098\cdot M \right\rceil$\\
    AC  & $240\, $ms & $\left\lfloor 0.6289\cdot M \right\rceil$\\
    SAC & $160\, $ms & $\left\lfloor 0.6289\cdot M \right\rceil$\\
    MoAPC  & $164\, $ms & $\left\lfloor 0.7736\cdot M \right\rceil$\\
    Huffman  & $229\, $ms & $\left\lfloor 0.6753\cdot M \right\rceil$\\
    
	MoPC$^{*}$ $\equiv$ MoHuffman \cite{lee2023isimitigating}  & $159\, $ms & $\left\lfloor 0.6753\cdot M \right\rceil$\\

	\hline
	\end{tabular} 
	\end{center}
	\renewcommand{\arraystretch}{1}
\vspace{12pt}
\end{table}

Average encoding length and 1-bit counts of all compared coding methods are given in Table III and V respectively for the Alphabets 1 and 2. The word length is chosen to be 20. For Alphabet 2, EOF symbol is not counted in the word length. That is, each word of Alphabet 2 comprises of 20 non-EOF symbols followed by the EOF symbol. For MoAPC, AC, and SAC, the encoding length and power consumption values have been computed by randomly choosing $10^6$ words from each corresponding alphabet. For other methods, these values have been  computed probabilistically.
\vspace{0.5pt}

In the simulation,  the signal interval and molecule count values are normalized based on the values of the Uncoded method. Accordingly, the normalized signal interval and molecule count per transmission of a 1-bit values for all the compared methods are given in the Table IV and VI for Alphabet 1 and 2, respectively. Note that the function, $\left\lfloor x \right\rceil$, rounds the given real number $x$ to the nearest integer.

\vspace{0.5pt}
We implemented our particle-tracking MC simulator based on the design of the simulator given in \cite{simulation}, which uses the distribution at (1). For simulation parameters, we have used the values given in Table VII\footnote{Parameters in Table VII are commonly used in MC literature, representing an MC channel where human insulin hormone is an information molecule \cite{ISI_mitigating_methods_2015}.}.
To use in the detection Algorithm 1, for each exemplary alphabet, we initially, on a set of 1024 random words of length 20, computed the optimal $spacing$, the optimal $a$, and the $min$\footnote{For ISI-Mitigating codes, as the existence of a 1-bit is guaranteed at each block \cite{best_channel_coding_2020}, the $min$ value is not computed. For others, to estimate the smallest possible $min$ value (calculated by taking the minimum value among all the number of absorbed molecules during each signal interval that corresponds to a 1-bit in the pilot signals), we scaled each calculated $min$ by a factor of $\frac{5}{6}$.} values for each respective method, at each different molecule count. In the simulation, using these pre-determined values of coefficients $a$, $spacing$ and $min$ in Algorithm 1, for each method at each different molecule count, we sent 5120 randomly chosen words of length 20.

\begin{table}[t]
	\begin{center}
	\caption{Average Encoding Length and Power Consumption for Alphabet 2 with Word Length 20}
	\renewcommand{\arraystretch}{1.1}
	\label{tbl_system_parameters}
	\begin{tabular}{p{2.8cm}@{\hskip 0.2cm}c@{\hskip 0.5cm}c}
	\hline
	\bfseries{Coding Method} & {\centering \bfseries{Encoding Length}} & \parbox[c]{2.5cm}{\centering \bfseries{Power Consumption} \\ \footnotesize (Number of 1-bits)} \\  
	\hline
	Uncoded & $$63$$ & $$16.73684$$ \\
 ISI-Mitigating \cite{best_channel_coding_2020} & $$105$$ & $$22$$ \\
 AC  & $$46.83747$$ & $$23.15221$$ \\
 SAC & $$69.98969$$ & $$23.15221$$ \\
 MoAPC  & $$68.72786$$ & $$18.61808$$ \\
 Huffman  & $$47.84210$$ & $$22.57894$$ \\
	MoHuffman \cite{lee2023isimitigating}   & $70.42105$& $$22.57894$$ \\
 MoPC$^{*}$  & $$68.84210$$ & $$16.73684$$ \\

	\hline
	\end{tabular} 
	\end{center}
	\renewcommand{\arraystretch}{1}
\vspace{5pt}
\end{table}

\begin{table}[t]
	\begin{center}
	\caption{Signal Interval and Molecule Count Normalizations for Alphabet 2 with Word Length 20}
	\renewcommand{\arraystretch}{1.1}
	\label{tbl_system_parameters}
	\begin{tabular}{p{2.8cm}@{\hskip 0.2cm}c@{\hskip 0.5cm}c}
	\hline
	\bfseries{Coding Method} & \parbox[c]{2cm}{\centering \bfseries{Signal Interval}} & \parbox[c]{2.5cm}{\centering \bfseries{Molecule Count} \\ \footnotesize ($M=100, 200,...$)} \\ 
	\hline
    Uncoded & $200\, $ms & $1\cdot M$ \\
    ISI-Mitigating \cite{best_channel_coding_2020} & $120\, $ms & $\left\lfloor 0.7607\cdot M \right\rceil$\\
    AC  & $269\, $ms & $\left\lfloor 0.7229\cdot M \right\rceil$\\
    SAC & $180\, $ms & $\left\lfloor 0.7229\cdot M \right\rceil$\\
    MoAPC  & $183\, $ms & $\left\lfloor 0.8989\cdot M \right\rceil$\\
    Huffman  & $263\, $ms & $\left\lfloor 0.7412\cdot M \right\rceil$\\
	MoHuffman \cite{lee2023isimitigating} & $179\, $ms & $\left\lfloor 0.7412\cdot M \right\rceil$\\
    MoPC$^{*}$  & $183\, $ms & $1\cdot M$\\
	\hline
	\end{tabular} 
	\end{center}
	\renewcommand{\arraystretch}{1}
\vspace{2pt}
\end{table}

Word error rate (WER) is defined as the ratio of the number of decoded words that do not perfectly match their corresponding original words to the total number of transmitted words. The simulation results are shown in Figs. 21, 22, 23, and 24, giving the respective WER and SER values. For a fair comparison among EOF-excluded methods, it is assumed that the receiver knows where the encodings end for all the transmissions of Alphabet 1. The simulation results show that the proposed MoAPC consistently outperformed SAC, which in turn always outperformed AC. These findings demonstrate that we have successfully adapted arithmetic source coding to molecular communication with significantly better channel reliability. For Alphabet 1, the proposed MoAPC achieved the best WER performance, and asymptotically outperformed all others in terms of SER. For Alphabet 2, MoPC$^{*}$ surpassed all other methods, including its main competitors MoHuffman and Huffman, in terms of both SER and WER.

\begin{figure}[t]
\centering
\includegraphics[width=0.8\linewidth]{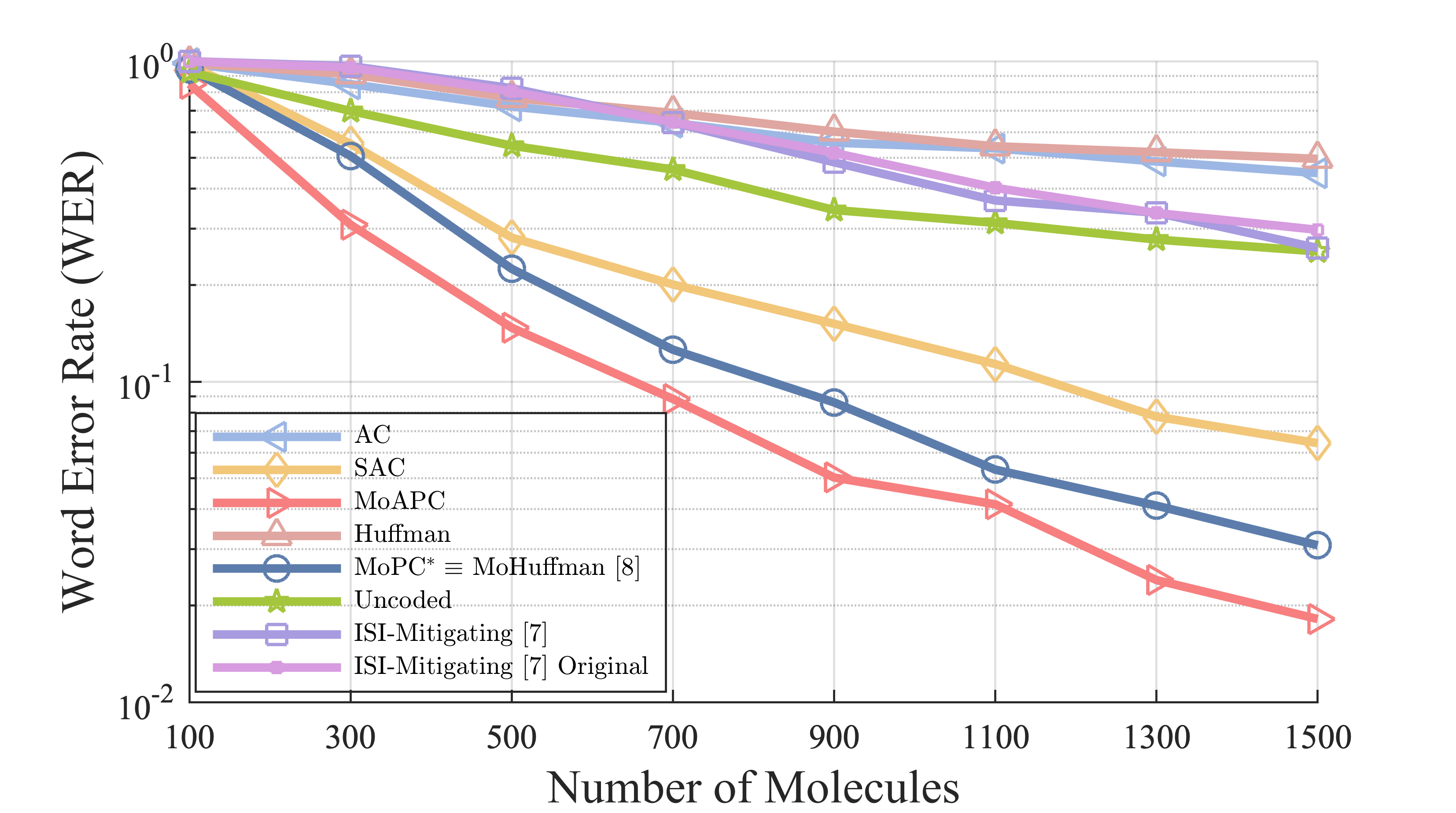}
\caption{Word Error Rates of Exemplary Alphabet 1}  
\includegraphics[width=0.8\linewidth]{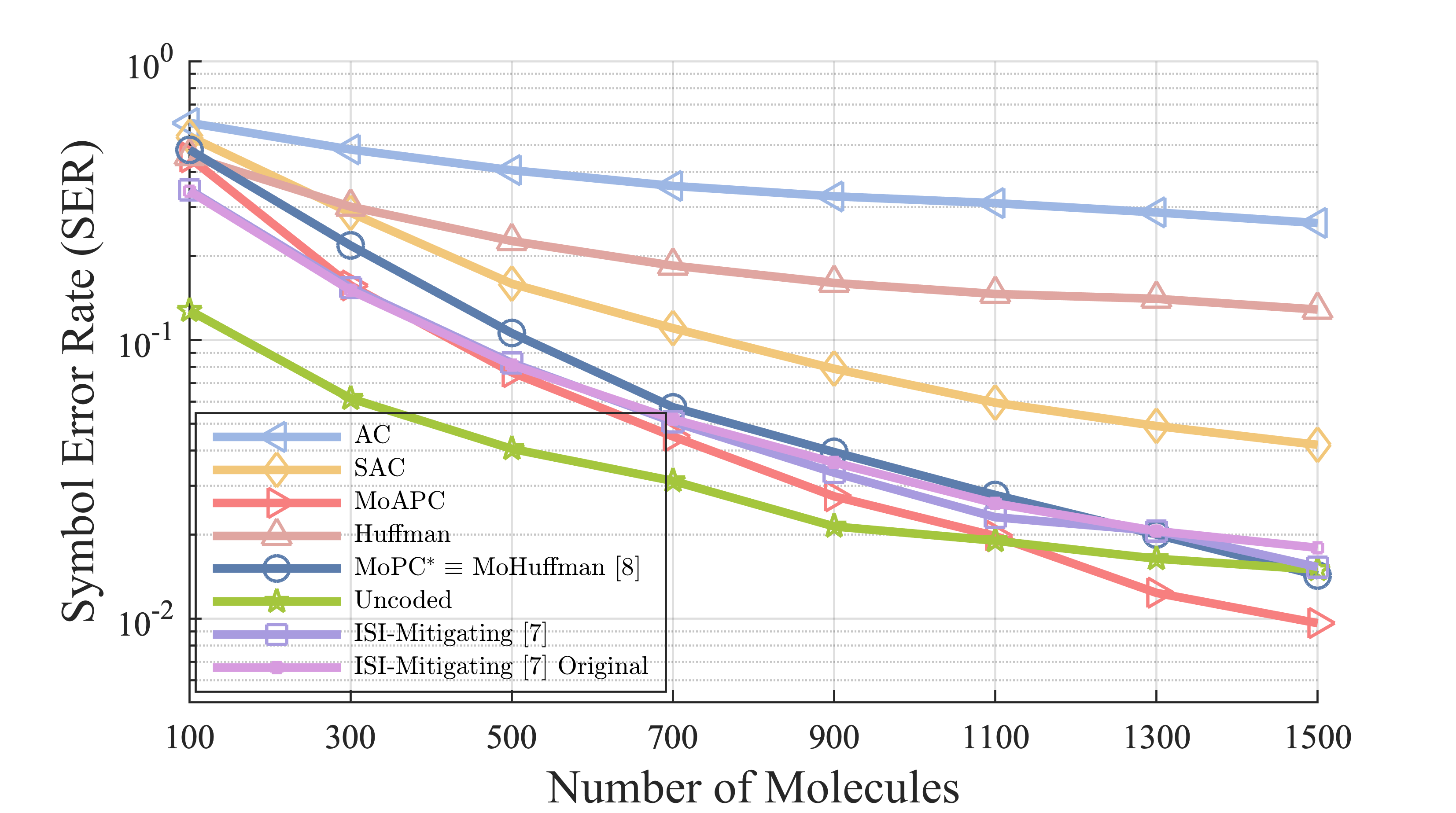}
\caption{ Symbol Error Rates of Exemplary Alphabet 1 }

\end{figure}
\begin{table}[t]
    \begin{center}
    \caption{Parameters Used in the Simulation}
    \renewcommand{\arraystretch}{1.1}
    \label{tbl_system_parameters}
    \begin{tabular}{p{5.5cm} l}
    \hline
    \bfseries{Parameter} 							& \bfseries{Value} \\ 
    \hline
    Diffusion coefficient ($D$) 	& $79.4\,\,\text{µm}^2/\text{s}$\\
    Distances between Tx and Rx ($r_0$)			& $10\,\, \text{µm}$\\
    Receiver radius ($r_R$)			& $5\,\,\text{µm}$\\
    Gaussian Counting Noise Variance ($\sigma_n^2$)			& 0\\
    Uncoded Signal Interval ($t_s$)			& $200\,\,\text{ms}$\\
    Particle-Tracking Simulator Step Size ($\Delta t$)			& $1\,\,\text{ms}$\\
    \hline
    \end{tabular} 
    \end{center}
    \renewcommand{\arraystretch}{1}
\vspace{7pt}
\end{table}

\vspace{2pt}
\vspace{-0.5cm}
\subsection{Self-Synchronisation Property and Future Work on MoAC} 

While arithmetic coding methods offer superior compression performance compared to prefix coding techniques, they lack the crucial property of self-synchronization. Self-synchronization allows a decoder to recover from bit errors after a certain number of symbols, ensuring more reliable decoding. In prefix coding, most codebooks possess this property \cite{self-correction}. Conversely, a single symbol error in arithmetic coding causes all subsequent symbols to be detected randomly based on the symbol distribution of the alphabet. 

Unless MoAPC has a significantly better compression and power consumption advantage over MoPC$^{*}$, as in Alphabet 1, it can be expected to perform inferiorly than MoPC$^{*}$ in highly stochastic MC channels, due to its lack of self-synchronization property. Several techniques can integrate self-synchronization into arithmetic coding. Soft decoding for error resilience is discussed in \cite{synchronization_of_arithmetic}. Marker methods for error detection are presented in \cite{marker1, marker2, marker3}, and error correction techniques are detailed in \cite{arithmetic-chase-like, arq_procedure}. Future research in MC source coding should prioritize the integration of these techniques into MoAC, equipping it with self-synchronization capabilities.

\subsection{Computational Considerations for MC Source Coding} 

Since MC is designed for nano-scale environments, circuit designs should remain relatively simple and efficient. Although finding a MoPC$^{*}$ prefix codebook is currently an exponential task, once found, it requires fewer computational resources than MoAC (and thus MoAPC) during encoding and decoding. However, for higher-order source coding, the number of prefix codebooks required increases exponentially with the data compression order \cite{data_compression_book}. As shown in Fig. 16 and 20, most of the encoding in MoAPC relies on MoAC. Therefore, since MoAC offers a superior compression performance compared to MoPC$^{*}$, the MoPC$^{*}$ component of MoAPC can be fixed at the zeroth order while still benefiting from the higher-order data compression advantages of MoAC. For higher order MC source coding, this approach could possibly make MoAPC a more effective choice than MoPC$^{*}$ by reducing the exponential space requirements associated with higher-order prefix coding.
\label{sec:simulations}

\vspace{-0.5cm}
\vspace{-1pt}
\section{Conclusion}

\vspace{-1pt}

\begin{figure}[t]
\centering
\includegraphics[width=0.8\linewidth]{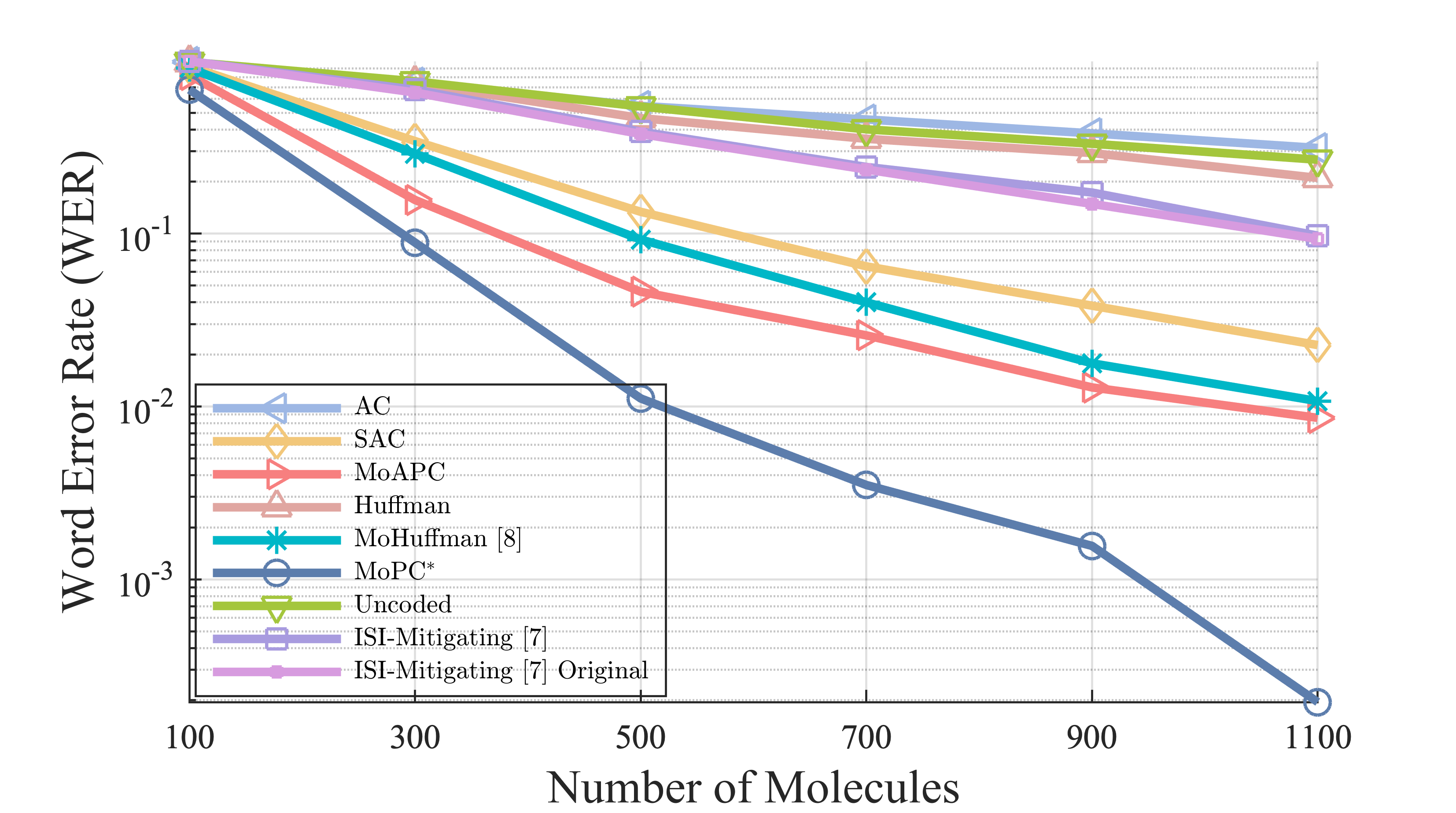}
\caption{Word Error Rates of Exemplary Alphabet 2}
\includegraphics[width=0.8\linewidth]{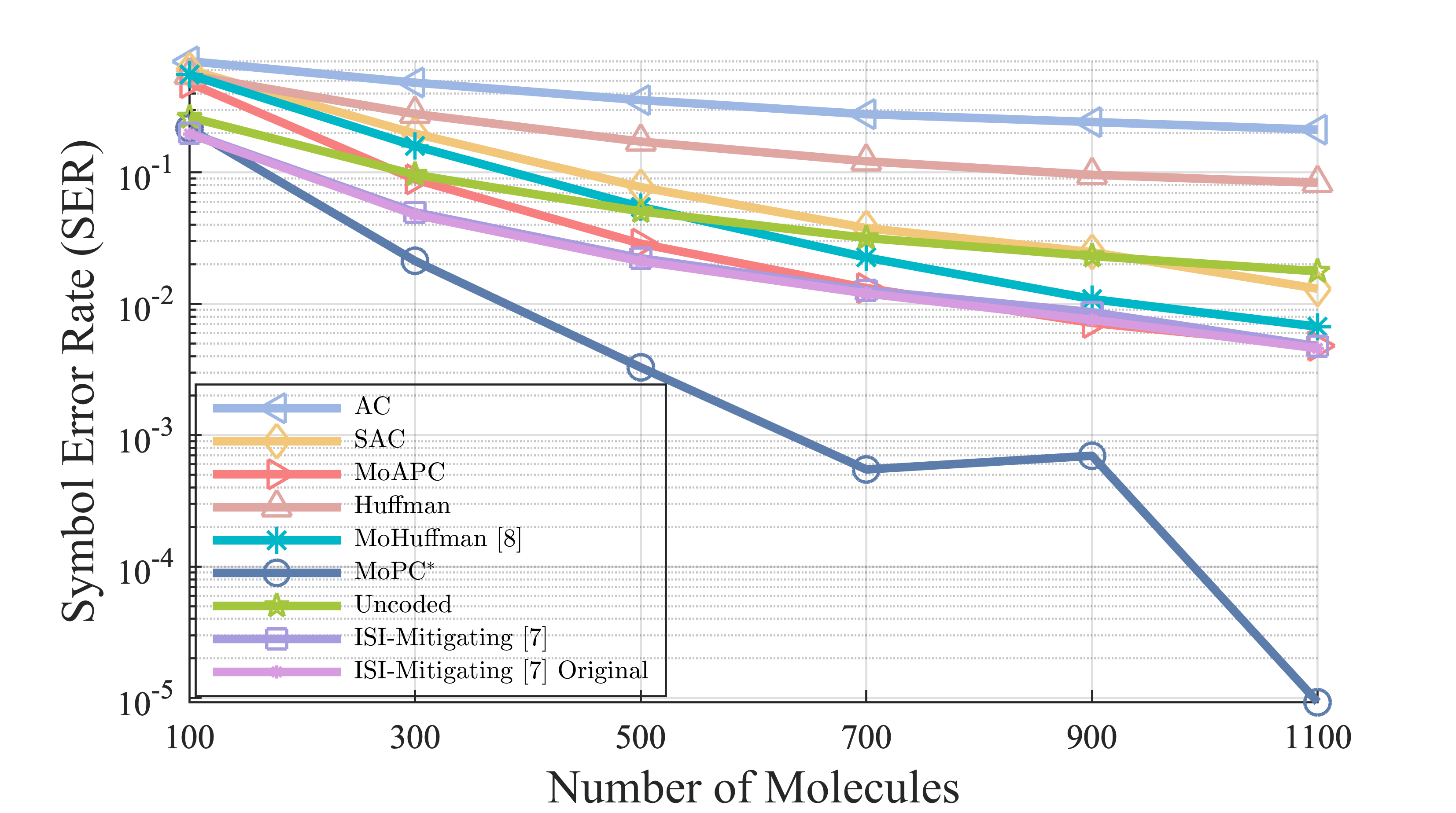}
\caption{Symbol Error Rates of Exemplary Alphabet 2}

\end{figure}

This paper proposes Molecular Arithmetic Coding (MoAC), a novel arithmetic source coding method to mitigate ISI in MC, which builds upon the general constrained arithmetic channel coding scheme in \cite{RLL_arithmetic}. To address the finite-precision limitations of MoAC, we have introduced Molecular Arithmetic with Prefix Coding (MoAPC), which guarantees unique decodability. Simulation results show that MoAPC outperforms both classical arithmetic coding (AC) and substitution arithmetic coding (SAC), which is our trivial adaptation of AC to MC. We have shown that MoHuffman \cite{lee2023isimitigating}, though it significantly improves channel reliability compared to conventional Huffman coding, does not always produce an optimal Molecular Prefix Coding (MoPC$^*$) codebook. Accordingly, we have used a brute-force algorithm in deriving MoPC$^*$ codebooks.

Future work should focus on developing polynomial-time algorithms for finding MoPC$^{*}$ codebooks. Additionally, integrating self-synchronization into MoAC can improve its reliability. In MC, accurate normalization of length and power values is essential. We have adopted the normalization procedure from \cite{normalization}, fitting it to the probabilistic nature of source coding. This normalization procedure should be adhered to in all future MC source coding research to ensure a fair comparison among different coding methods. The application areas of MoAC extend well beyond MC, addressing broader contexts (such as wireless sensor networks) where transmitting a 1-bit incurs higher energy costs than a 0-bit—a phenomenon known as energy consumption disparity (ECD) \cite{energy_consumption_disparity}. We proved that MoAC reduces the transmission of 1-bits approximately by 25\% compared to AC, leading to significant energy savings. 
\vspace{-1pt}
\vspace{6pt}
\vspace{-13pt}
\vspace{3pt}
\section*{Code Availability}
\vspace{-2pt}
Pseudocode for MoAC, as well as Python implementations of MoAC and AC (with and without EOF versions), can be found at: https://github.com/MelihSahinEdu/MCArithmetic.git

\vspace{1pt}
\vspace{-0.3cm}
\section*{Acknowledgment}
\vspace{-2pt}
 Melih Şahin, the first author, dedicates this work to his late grandfather Muzaffer Şahin, whose battle with cancer inspired him to begin research in Molecular Communication. This work is driven by the hope that advancements in Molecular Communication will lead to more effective cancer treatments.
\vspace{-0.7cm}

\bibliographystyle{IEEEtran}

\bibliography{References}
\vspace{-20pt}

\begin{IEEEbiography}[{\includegraphics[width=1in,height=1.25in,clip,keepaspectratio]{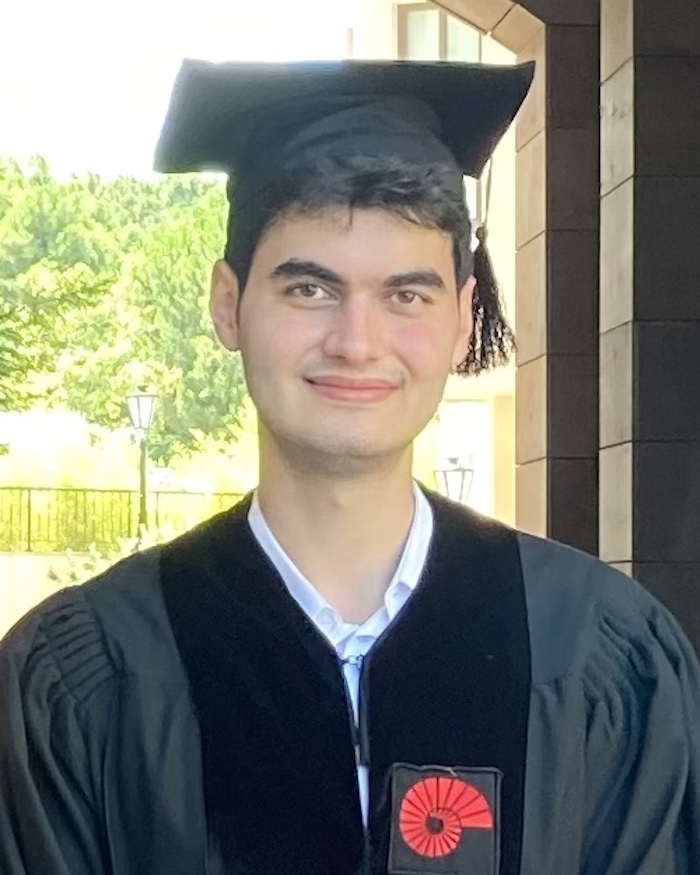}}]{Melih Şahin (Student Member, IEEE)}
graduated from both Ankara Science High School and Bilsem in 2019. He started studying a B.Sc. degree in School of Computing at Korea Advanced Institute of Science and Technology (KAIST) in 2020, completing first year of his degree in 2021. He then transferred to Koç University, Türkiye, where he graduated with a B.Sc. degree in Computer Engineering, together with a minor in Mathematics and a track on AI in June 2024. He is currently pursuing a MSc. degree in Electrical and Electronics Engineering at Koç University. He holds many prestigious international and national awards in scientific research and painting. His most notable achievements include a 4$^{th}$ place Grand Award in Mathematics at Intel International Science and Engineering Fair (ISEF) 2018 Pittsburgh, as well as a global bronze medal (3$^{rd}$ place) at Toyota International My Dream Car Art Contest 2013 in the age category of 10-12. His current research interests are Coding Theory, Molecular Communication, Quantum Computing, Computation Theory, Machine Learning, and Algebraic Combinatorics.

\end{IEEEbiography}

\vspace{-0.3cm}
\begin{IEEEbiography}[{\includegraphics[width=1in,height=1.25in,clip,keepaspectratio]{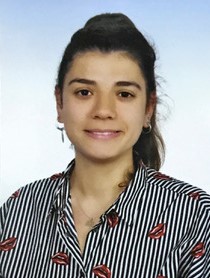}}]{Beyza Ezgi Ortlek (Student Member, IEEE)}
received her B.Sc. degree in Electrical and Electronics Engineering and in Molecular Biology and Genetics as a double major from Koç University, Istanbul, Türkiye, where she is currently pursuing a Ph.D. degree in Electrical and Electronics Engineering. She is a research assistant at the Next-generation and Wireless Communications Laboratory. Her research interests include Intrabody Nanonetworks and Molecular Communications.
\end{IEEEbiography}

\vspace{-0.3cm}
    \begin{IEEEbiography}[{\includegraphics[width=1in,height=1.25in,clip,keepaspectratio]{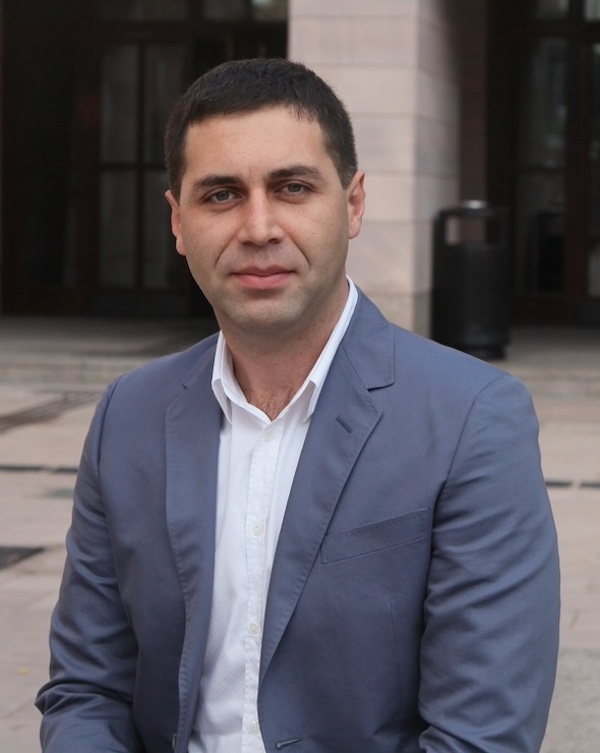}}]{Ozgur B. Akan (Fellow, IEEE)}

 received the PhD from the School of Electrical and Computer Engineering Georgia Institute of Technology Atlanta, in 2004. He is currently the Head of Internet of Everything (IoE) Group, with the Department of Engineering, University of Cambridge, UK and the Director of Centre for neXt-generation Communications (CXC), Koç University, Türkiye. His research interests include wireless, nano, and molecular communications and Internet of Everything.
\end{IEEEbiography}

\end{document}